\documentclass{article}
\usepackage{amsfonts}
\usepackage{amsmath,amssymb, amscd}
\usepackage{graphicx}
\usepackage{color}

\newtheorem{theorem}{Theorem}
\newtheorem{acknowledgement}[theorem]{Acknowledgement}

\newtheorem{lemma}[theorem]{Lemma}

\begin{document}

\title{Stability of selfsimilar solutions to the fragmentation equation with
polynomial daughter fragments distribution}
\author{Marco A. Fontelos \\
ICMAT-CSIC, C/Nicol\'{a}s Cabrera, no 13-15 Campus de Cantoblanco,\\
UAM, 28049 Madrid, Spain}
\maketitle

\begin{abstract}
We study fragmentation equations with power-law fragmentation rates and
polynomial daughter fragments distribution function $p(s)$. The
corresponding selfsimillar solutions are analysed and their exponentially
decaying asymptotic behaviour and $C^{\infty }$ regularity deduced.
Stability of selfsimilar solutions (under smooth exponentially decaying
perturbations), with sharp exponential decay rates in time are proved, as
well as $C^{\infty }$ regularity of solutions for $t>0$. The results are
based on explicit expansion in terms of generalized Laguerre polynomials and
the analysis of such expansions. For perturbations with power-law decay at
infinity stability is also proved. Finally, we consider real analytic $p(s)$.
\end{abstract}

\section{Introduction}

Coagulation-fragmentation theory is a very active area of research involving
theoretical disciplines like pure analysis, mathematical physics and
probability, and more practical fields among which we can include integral
equations, numerical analysis and stochastic processes (random graphs and
SPDEs). However, the origins of this theory lie in aerosol physics and a
variety of applications have risen even before the first rigorous formal
results could be established. Here we provide a short list, incomplete due
to the extensive development of applications, of representative fields of
application integrating references from \cite{Aldous99,LaurMisch04,Leyv03}
with some more recent study; further references can also be found in \cite%
{Drake72review,Wattis06,BellouBian10}. Fragmentation processes appear in
various scientific contexts such as: depolymerization \cite%
{ZiffMcGrady85,Ziff91}, erosion: \cite{Kostoglou01b}, sprays and drop
dispersion \cite{ClaLash99}; breakage of immiscible fluids in turbulent
flows \cite{EastwArmiLash04,LasherasEtAll02}; fragmentation-coagulation
scattering model for the dynamics of stirred liquid-liquid dispersions \cite%
{FasanoEtAll06}, economy and social systems: \cite%
{AjmBellomoEgidi08,Ajmone09,ZhaoEtAll09}, cell cultures, multicellular
growing systems and biological tissues \cite{BellomoEtAll2009}.

As one can deduce from the literature, the mathematical theory of
fragmentation is far more developed than its coagulation counterpart (a
general review is provided in \cite{B2}), evidently due to the fact that the
theory of linear operators gives general and stronger results. Concerning
the study of the \emph{self-similar fragmentation equation}, previous work
can be found in the following references: Cheng and Redner \cite{ChengRed90}
give a first general discussion of the kinetics of continuous, irreversible
fragmentation processes; Treat in \cite{Treat97a} and mainly in \cite%
{Treat97b} settles some fundamental facts about the similarity solutions;
Bertoin \cite{Bertoin06} develops the probabilistic counterpart
(fragmentation stochastic processes); Escobedo, Mischler and
Rodriguez-Ricard \cite{EscoMischRic05}, notably, completely solve the
existence of self-similar solutions and the convergence problem; later,
Michel \cite{Michel06,Michel06b} deals with the cell division eigenproblem;
Lauren\c{c}ot and Perthame \cite{LaurBen09} develop bounds for the
exponential decay for the growth-fragmentation or cell-division equation;
these problems (in particular the growth-fragmentation equations) and the
rate of convergence are later considered by C\'{a}ceres, Ca\~{n}izo and
Mischler \cite{CaceCaMisc11} and Balagu\'{e}, Ca\~{n}izo and Gabriel \cite%
{BalaCaGabr13}. The study of fragmentation equations from an evolution
semigroup point of view, has also been the object of many studies (see \cite%
{AB} and references therein). A general and detailed account of the theory
of fragmentation processes is contained in the two volumes \cite{B1} and
\cite{B2}.

The fragmentation equation is

\begin{equation}
\frac{\partial f}{\partial t}(x,t)=\int_{x}^{\infty
}b(y,x)f(y,t)dy-B(x)f(x,t),  \label{fragmentation}
\end{equation}%
to be solved \ in $x\in
\mathbb{R}
^{+}$ and $t>0$ with initial data $f(x,0)=f_{0}(x)$. The function $B(x)$ is
the fragmentation rate, that is the rate at which elements of size $x$
undergo fragmentation. The distribution of fragment sizes $y$ produced when
an element of size $x$ breaks will be determined by $b(y,x)$. We will take%
\begin{equation}
B(x)=x^{\gamma },\ b(y,x)=y^{\gamma -1}p\left( \frac{x}{y}\right) ,
\label{cond1}
\end{equation}%
where $\gamma >0$ and $p(s)$ is a function defined in the interval $\left[
0,1\right] $ and representing the daughter fragments distribution. The fact
that the total mass or size of daughters produced in a fragmentation has to
equal the initial mass or size implies the condition%
\begin{equation*}
\int_{0}^{1}sp(s)ds=1.
\end{equation*}%
Of special importance are the selfsimilar solutions of (\ref{fragmentation})
as the best candidates to represent the long term distribution of particle
sizes. Selfsimilar solutions are known to represent the long time
asymptotics for multiple linear and nonlinear problems, as well as in
nonlinear partial differential equations where finite-time singularities
ocurr (see \cite{EF} for a general review). The selfsimilar solutions are of
the form%
\begin{equation*}
f(x,t)=(1+t)^{\frac{2}{\gamma }}u_{s}(x(1+t)^{\frac{1}{\gamma }})
\end{equation*}%
or, equivalently, stationary solutions of the equation
\begin{equation}
\frac{\partial u}{\partial t}+\frac{\partial }{\partial x}(xu)+u=\gamma
\left( \int_{x}^{\infty }b(y,x)u(y)dy-B(x)u(x)\right) ,  \label{eqn}
\end{equation}%
where the function $u$ is related to $f$ by means of the following
similarity change of variables:%
\begin{equation*}
f(x,t)=(1+t)^{\frac{2}{\gamma }}u(\gamma ^{-1}\log (1+t),x(1+t)^{\frac{1}{%
\gamma }}).
\end{equation*}%
Given the linearity of equation (\ref{eqn}), the selfsimilar solution $%
u_{S}(y)$ can be chosen such that its total mass (first moment) coincides
with that of a given initial data $u(x,0)=u_{0}(x)$ to the evolution problem:%
\begin{equation*}
\int_{0}^{\infty }xu_{S}(x)dx=\int_{0}^{\infty }xu_{0}(x)dx.
\end{equation*}%
Then, by writing $u_{0}(x)=u_{S}(x)+u^{\ast }(x,t)$, one finds that $u^{\ast
}(x,t)$ satisfies (\ref{eqn}) and is such that%
\begin{equation*}
\int_{0}^{\infty }xu^{\ast }(x,t)dx=0,
\end{equation*}%
so that stability of selfsimilar solutions implies studying the evolution
problem (\ref{eqn}) in sets of functions with zero mass (and non-positive
therefore).

The existence of selfsimilar functions, and their properties, for various
functions $p(s)$ have been addressed in multiple publications. Seminal works
focused in polynomial functions $p(s)$ (cf. \cite{Treat97a} and \cite%
{Treat97b}) while more recent works consider other $p(s)$ such as a Dirac
delta centered at $s=\frac{1}{2}$ (representing a mitosis process),
combination of Dirac deltas, functions compactly supported away from zero,
etc (see \cite{EscoMischRic05}, \cite{BreschiFontelos} and references
therein). Stability of the selfsimilar solutions has not been so widely
studied, with $p(s)$ constant or close to a constant (see \cite{CaceCaMisc11}
for the case $\gamma \leq 2$ and \cite{GabrielSalvarani} for $\gamma >2$,
see also \cite{MS} for a more abstract but also more general study of
stability of selfsimilar solutions by means of semigroup theory). In these
works, the main result is a spectral gap theorem stating that the distance
in a suitable norm between the solution of (\ref{eqn}) and the selfsimilar
solution with the same mass as the initial data, decays exponentially fast.
These results are not sharp in the exponent and do not trivially extend to
more complex functions $p(s)$. In this paper we will provide explicit
representations for the solution of (\ref{eqn}) \ that will allow us to
deduce sharp decay rates $O(e^{-\gamma t})$ if initial data are smooth and
present exponential decay at infinity. The results extend to any polynomial
and any positive $\gamma $ under suitable conditions on the initial data.
For initial data with power-law decay at infinity, the decay in time depends
on $\gamma $. If $\gamma <1$, the decay is still $O(e^{-\gamma t})$ but, for
$\gamma \geq 1$, the decay is $O(e^{-t})$.

The article is organized as follows. In Section 2 we introduce the main
tools, mainly based on Mellin transform techniques, to be used in the next
sections and present the main results in the form of four theorems. In
section 3 we study the selfsimilar solutions, while sections 4 and 5 are
concerned with stability results in two different functional spaces: section
4 is for functions decreasing exponentially fast at infinity while section 5
is for $L^{2}$ spaces with power-like weights. In section 6 we discuss some
extensions to more general $p(s)$ and also describe the limitations of the
methods developed. In particular, we extend results of previous sections to $%
p(s)$ real analytic under some additional conditions. Finally, a first
appendix provides the Mellin transform of functions used along the text, the
second appendix is for properties involving Gamma functions and their
product while the last appendix provides a proof of a representation formula
for solutions in the case of constant $p(s)$.

\section{Preliminaries and main results}

In this section we will present the main tools to be used in the next
sections. Firstly, it will be critical the use of the so-called Mellin
transform (of a function $U(x)$) defined by

\begin{equation}
\widetilde{U}(z)=\int_{0}^{\infty }x^{z-1}U(x)dx,  \label{m}
\end{equation}%
where $z\in
\mathbb{C}
$. Throughout this text we will also denote Mellin transforms of a function $%
U(x)$ as $\widetilde{U}(z)$. The corresponding inverse Mellin transform is
then given by
\begin{equation}
U(x)=\frac{1}{2\pi i}\int_{-i\infty +\delta }^{i\infty +\delta }x^{-z}%
\widetilde{U}(z)dz,  \label{invm}
\end{equation}%
where the integration is done along the line $\Re (z)=\delta $ in the
complex plane. The parameter $\delta $ is taken so that $0<\delta \ll 1$ and
possible poles of $\widetilde{U}(z)$ at $\Re (z)=0$ lay at the left of the
integration contour. Notice that%
\begin{equation*}
\int_{0}^{\infty }x^{z-1}U(x)dx=\int_{0}^{\infty }e^{z\log x}U(x)d\log
x=\int_{-\infty }^{\infty }e^{zy}U(e^{y})dy,
\end{equation*}%
so that $\widetilde{U}(z)$ may be viewed as the Fourier transform of $%
U(e^{y})$. One can then prove the following Plancherel formula:%
\begin{equation}
\int_{0}^{\infty }\left\vert U(x)\right\vert ^{2}dx=\frac{1}{2\pi }%
\int_{-\infty }^{\infty }\left\vert \widetilde{U}(i\lambda +\frac{1}{2}%
)\right\vert ^{2}d\lambda .  \label{plan}
\end{equation}

If we consider now the following integral transform of $g(x)$:
\begin{equation*}
h(x)=\int_{0}^{\infty }f\left( x/y\right) y^{-1}g(y)dy,
\end{equation*}%
we compute its Mellin transform as%
\begin{eqnarray}
H(z) &=&\int_{0}^{\infty }x^{z-1}h(x)dx  \notag \\
&=&\int_{0}^{\infty }x^{z-1}\left( \int_{0}^{\infty }f\left( x/y\right)
y^{-1}g(y)dy\right) dx  \notag \\
&=&\int_{0}^{\infty }\int_{0}^{\infty }s^{z-1}y^{z-1}f\left( s\right)
g(y)dyds=F(z)G(z),  \label{p0}
\end{eqnarray}%
which is a relation similar to that of the Fourier transform of a
convolution. In the Appendix A we provide the Mellin transforms of funtions
that will be used along the text.

Coming back to equation (\ref{eqn}) we will, for the sake of convenience,
change variables $x\rightarrow x^{\prime }=x^{\gamma }$ so that
\begin{equation}
\frac{\partial u}{\partial t}+\gamma x\frac{\partial u}{\partial x}%
+2u=\gamma \left( \frac{1}{\gamma }\int_{x}^{\infty }u(y)p\left( \left(
\frac{x}{y}\right) ^{\frac{1}{\gamma }}\right) dy-xu(z)\right) ,
\label{eqnt}
\end{equation}%
(after renaming $x^{\prime }$ as $x$) and perform Mellin transform
\begin{equation*}
\widetilde{u}(z,t)=\int_{0}^{\infty }x^{z-1}u(x,t)dx,
\end{equation*}%
to obtain the equation%
\begin{equation}
\widetilde{u}_{t}(z,t)-\gamma z\widetilde{u}(z,t)+2\widetilde{u}(z,t)=\gamma
\left( \frac{1}{\gamma }P(z)\widetilde{u}(z+1,t)-\widetilde{u}(z+1,t)\right)
,  \label{eqnMellin}
\end{equation}%
where, being $p(s)=\sum_{i=0}^{N}a_{i}s^{i}$, one has%
\begin{equation*}
P(z)=\int_{0}^{1}s^{z-1}p(s^{\frac{1}{\gamma }})ds=\sum_{i=0}^{N}\frac{a_{i}%
}{z+i/\gamma }.
\end{equation*}%
Note that equation (\ref{eqnMellin}) does not contain derivatives but,
instead, relates $\widetilde{u}$ at two different points $z$ and $z+1$ in
the complex plane, making it evident the intrinsic nonlocality of the
problem. Nevertheless, this formulation will allow to obtain an almost
explicit solution to the problem. The right hand side of (\ref{eqnMellin})
may be written as $\widetilde{u}(z+1,t)$ multiplied by the multiplier
\begin{equation}
K(z)=\gamma \left( \frac{1}{\gamma }P(z)-1\right) ,  \label{k}
\end{equation}%
that can also be written as%
\begin{equation}
K(z)=-\gamma \frac{z-\frac{2}{\gamma }}{z}\frac{P_{N}(z)}{Q_{N}(z)},
\label{fact}
\end{equation}%
where%
\begin{eqnarray*}
P_{N}(z) &=&\prod_{i=1}^{N}(z-z_{i}), \\
Q_{N}(z) &=&\prod_{i=1}^{N}\left( z+\frac{i}{\gamma }\right) ,
\end{eqnarray*}%
with $\left\{ z_{i}\right\} $ are the $N$ complex roots (counted with
possible multiplicities) of an $N$-degree polynomial. In this paper we will
make the following assumption on the roots of $K(z)$:%
\begin{equation}
\text{The only root of }K(z)\text{ defined by (\ref{k}) in }\Re (z)>0\text{
is }z=\frac{2}{\gamma }.  \label{assumpt}
\end{equation}%
Assumption (\ref{assumpt}) can be shown to hold for any polynomial $p(s)$
with $a_{0}>0$ up to degree 3 and for higher degree polynomials under very
general conditions.

By expanding for large values of $z$%
\begin{eqnarray*}
\left( \frac{1}{\gamma }P(z)-1\right) &=&\left( -\frac{1}{\gamma }%
\sum_{i=0}^{N}\frac{a_{i}}{z+i/\gamma }+1\right) =\left( 1-\sum_{i=0}^{N}%
\frac{a_{i}}{\gamma z}+\sum_{i=0}^{N}\frac{ia_{i}}{\gamma ^{2}z^{2}}%
+...\right) \\
&=&\frac{1}{z}\left( z-\frac{2}{\gamma }\right) +\left( \frac{2}{\gamma }%
-\sum_{i=0}^{N}\frac{a_{i}}{\gamma }\right) \frac{1}{z}+O(z^{-2}),
\end{eqnarray*}%
and%
\begin{eqnarray*}
\frac{1}{z}\left( z-\frac{2}{\gamma }\right) \frac{\prod (z-z_{i})}{\prod (z+%
\frac{i}{\gamma })} &=&\frac{1}{z}\left( z-\frac{2}{\gamma }\right) \frac{%
\prod (1-\frac{z_{i}}{z})}{\prod (1+\frac{i}{\gamma z})} \\
&=&\frac{1}{z}\left( z-\frac{2}{\gamma }\right) \frac{(1-\sum \frac{z_{i}}{z}%
+...)}{(1+\sum \frac{i}{\gamma z}+...)} \\
&=&\frac{1}{z}\left( z-\frac{2}{\gamma }\right) -\frac{1}{z}\sum (i/\gamma
+z_{i})+O(z^{-2}),
\end{eqnarray*}%
we can prove the following relation between the coefficients $\left\{
a_{i}\right\} _{i=0}^{N}$ and the roots $\left\{ z_{i}\right\} _{i=1}^{N}$:%
\begin{equation}
\nu \equiv \sum_{i=1}^{N}(i/\gamma +z_{i})=\sum_{i=0}^{N}\frac{a_{i}}{\gamma
}-\frac{2}{\gamma }.  \label{nu}
\end{equation}

The parameter $\nu $, for reasons that will be apparent later, will be
called the regularity index. Note that the regularity index is zero if $%
p(s)=2$, is positive if $\sum_{i=0}^{N}a_{i}=p(1)>2$ and negative if $p(1)<2$%
. In particular, if $p(s)$ is an increasing function, then $p(1)>2$ and the
regularity index is positive.

The main idea behind our approach is to factor $\widetilde{u}(z,t)$ in the
form
\begin{equation*}
\widetilde{u}(z,t)=R(z)\Psi (z,t),
\end{equation*}%
where $R(z)$ satisfies the functional equation%
\begin{equation}
R(z)=\frac{P_{N}(z)}{Q_{N}(z)}R(z+1),  \label{func}
\end{equation}%
while $\Psi (z,t)$ satisfies the equation
\begin{equation}
\Psi _{t}(z,t)=\gamma \left( z-\frac{2}{\gamma }\right) \left[ \Psi (z,t)-%
\frac{1}{z}\Psi (z+1,t)\right] ,  \label{eqpsi}
\end{equation}%
which is independent of the distribution $p(s)$ and is, in fact, the
equation to be satisfied when $p(s)=2$. The simplicity of (\ref{eqpsi}) will
allow us to solve it in an almost explicit manner. The functional equation (%
\ref{func}) involves the quotient of two $N$-th degree polynomials, which is
a meromorphic function and we will be able to solve it in terms of gamma
functions whose asymptotic behaviors are simple to estimate rendering a
multiplier operator with the characteristics of a Fourier multiplier
corresponding to a fractional integral/derivative of order $\nu $. This will
allow to obtain estimates of $\widetilde{u}$ in terms of suitable norms of $%
\Psi $ and relate these norms (by (\ref{plan})) with classical Lebesgue and
Sobolev norms.

This approach through Mellin's transform allows us to prove, in the next
sections, the following four theorems concerning the equation (\ref{eqn})
with (\ref{cond1}). We will assume, throughout this paper, that $a_{0}>0$
and (\ref{assumpt}) holds. The degenerate case $a_{0}=0$ will be discussed
in the final section. The first theorem concerns selfsimilar solutions to
fragmentation equations, i.e. stationary solutions of (\ref{eqn}) with (\ref%
{cond1}).

\begin{theorem}
\label{th1}There exists a finite mass and nonnegative stationary solution $%
u_{S}(x)$ to (\ref{eqn}) with (\ref{cond1}) and $p(s)$ a polynomial of
degree $N$ (with $a_{0}>0$). The solution $u_{S}(x)\in C^{\infty }(0,\infty
) $ and is such that%
\begin{equation*}
u_{S}(x)\sim x^{p(1)-2}e^{-x^{\gamma }}(1+O(x^{-\gamma })),\ \ \text{as }%
x\rightarrow \infty .
\end{equation*}
\end{theorem}

\bigskip The second theorem concerns the evolution problem (\ref{eqn}) with (%
\ref{cond1}) and $p(s)=2$. By the results in section 5.1.3.4 of \cite{B1},
the solution to (\ref{fragmentation}) is unique in the class of positive and
mass preserving functions. Since the selfsimilar solutions are nonnegative
and have finite mass, it is sufficient to consider zero mass initial
conditions $u_{0}(x)$ representing perturbations to $u_{S}(x)$ (and such
that $u_{0}+u_{S}$ is nonnegative). We take as initial data $u_{0}(x)$ such
that%
\begin{equation}
\int_{0}^{\infty }x^{\gamma -1}e^{x^{\gamma }}\left\vert \left( x\frac{d}{dx}%
\right) ^{j}u_{0}(x)\right\vert ^{2}dx<C(j!)^{2(1+\sigma )}\text{ and }%
\int_{0}^{\infty }xu_{0}(x)dx=0.  \label{cond2}
\end{equation}%
for $j=0,1,\ldots $ and some $\sigma >0$. \ The first condition in (\ref%
{cond2}) implies that $u_{0}(x)$ belongs to the weighted Sobolev-Gevrey
class $\mathcal{G}_{1+\sigma }$.

\begin{theorem}
\label{th2}Let $u_{0}(x)$ satisfy (\ref{cond2}). Then the solution $u(x,t)$
to (\ref{eqn}) with (\ref{cond1}) and $p(s)=2$ is given by the formula%
\begin{eqnarray}
u(x,t) &=&\sum_{n=1}^{\infty }a_{n}e^{-n\gamma t}n!L_{n}^{\left( \frac{2}{%
\gamma }-n\right) }(x^{\gamma })e^{-x^{\gamma }}  \label{series} \\
a_{n} &=&\sum_{j=n}^{\infty }\frac{j!}{(n!)^{2}(j-n)!}\frac{\gamma }{\Gamma (%
\left[ \frac{2}{\gamma }\right] +1)}\int_{0}^{\infty }y^{\gamma
+1}L_{j}^{\left( \left[ \frac{2}{\gamma }\right] \right) }(y^{\gamma
})v_{0}(y)dy,  \notag
\end{eqnarray}%
with%
\begin{eqnarray*}
v_{0}(x) &=&\frac{1}{x^{\gamma \delta }\Gamma (\delta )}\left(
\int_{x}^{\infty }(1-\left( x/y\right) ^{\gamma })^{\delta -1}y^{\gamma
\delta -1}u_{0}(y)dy\right) \text{ if }\delta =\frac{2}{\gamma }-\left[
\frac{2}{\gamma }\right] >0 \\
v_{0}(x) &=&u_{0}(x)\ \text{if }\delta =0.
\end{eqnarray*}%
The following estimate holds true for $t>0$, $x>0$ and $\alpha >0$:%
\begin{equation*}
\left\vert u(x,t)\right\vert \leq C_{\alpha }e^{-\gamma t}e^{-(1-\alpha
)x^{\gamma }},
\end{equation*}%
as well as%
\begin{equation}
\left\vert \frac{\partial ^{k}u}{\partial (x^{\gamma })^{k}}(x,t)\right\vert
\leq C_{k,\alpha }e^{-\gamma t}e^{-(1-\alpha )x^{\gamma }},  \label{smooth}
\end{equation}%
for any $k\in
\mathbb{N}
$.
\end{theorem}

We remark that each term in the series (\ref{series}) carries zero mass
since, by Rodrigues formula and integration by parts,%
\begin{eqnarray*}
\int_{0}^{\infty }xL_{n}^{\left( \frac{2}{\gamma }-n\right) }(x^{\gamma
})e^{-x^{\gamma }}dx &=&\gamma ^{-1}\int_{0}^{\infty }y^{\frac{2}{\gamma }%
-1}L_{n}^{\left( \frac{2}{\gamma }-n\right) }(y)e^{-y}dy \\
&=&\gamma ^{-1}\int_{0}^{\infty }\frac{y^{n-1}}{n!}\frac{d^{n}}{dy^{n}}(y^{%
\frac{2}{\gamma }}e^{-y})dy \\
&=&(-1)^{n}\gamma ^{-1}\int_{0}^{\infty }\frac{1}{n!}\frac{d^{n}y^{n-1}}{%
dy^{n}}(y^{\frac{2}{\gamma }}e^{-y})dy=0.
\end{eqnarray*}%
Also note that (\ref{smooth}) implies $u(x,t)\in C^{\infty }(0,\infty )$ for
any $t>0$.

The third theorem concerns a general polynomial $p(s)$ of degree $N$. The
requirements of the initial data $u_{0}(x)$ are the also (\ref{cond2}).

\begin{theorem}
\label{th3}Let $u_{0}(x)$ satisfy (\ref{cond2}). Let $u(x,t)$ be a solution
to (\ref{eqn}) with (\ref{cond1}) and $p(s)$ a polynomial of degree $N$
(with $a_{0}>0$). Then%
\begin{equation*}
\left\vert u(x,t)\right\vert \leq Cx^{-\sigma }e^{-\gamma t}e^{-\frac{1}{2}%
x^{\gamma }},
\end{equation*}%
for $t>0$, $x>0$ and arbitrarily small $\sigma >0$. Moreover,%
\begin{equation}
\left\vert \left( x\frac{\partial }{\partial x}\right) ^{k}u(x,t)\right\vert
\leq C_{k}x^{-\sigma }e^{-\gamma t}e^{-\frac{1}{2}x^{\gamma }},  \label{ufp}
\end{equation}%
for any $k\in
\mathbb{N}
$.
\end{theorem}

We remark that estimate (\ref{ufp}) implies that $u(x,t)\in C^{\infty
}(0,\infty )$ for any $t>0$.

Finally, we will consider the evolution problem in a broader functional
setting $L_{1}^{2}(%
\mathbb{R}
^{+})=\left\{ u(x):\left\Vert u\right\Vert _{L_{1}^{2}(%
\mathbb{R}
^{+})}^{2}\equiv \int x\left\vert u(x)\right\vert ^{2}dx<\infty \right\} $
and also with the condition $\int_{0}^{\infty }xu_{0}(x)dx=0$ . This allows
for initial data with power-law decay at infinity. \ Analogously, we define
the space
\begin{equation*}
H_{1}^{\nu }(%
\mathbb{R}
^{+})=\left\{ u(x):\left\Vert u\right\Vert _{H_{1}^{\nu }(%
\mathbb{R}
^{+})}^{2}\equiv \int x\left\vert D_{\nu }u(x)\right\vert ^{2}dx<\infty
\right\}
\end{equation*}
where%
\begin{equation*}
\widetilde{D_{\nu }u}(z)=(1+\left\vert z\right\vert ^{2})^{\frac{\nu }{2}}%
\widetilde{u}(z),
\end{equation*}%
and note that, for $\nu =n\in
\mathbb{N}
$,
\begin{equation*}
\left\Vert u\right\Vert _{H_{1}^{\nu }(%
\mathbb{R}
^{+})}^{2}\sim \int_{0}^{\infty }x\left( u^{2}(x)+x^{2\nu }\left( \frac{%
d^{\nu }}{dx^{\nu }}u(x)\right) ^{2}\right) dx.
\end{equation*}%
We will prove the following Theorem:

\begin{theorem}
\label{th4}Let $u_{0}(x)\in L_{1}^{2}(%
\mathbb{R}
^{+})\cap L_{1}^{1}(%
\mathbb{R}
^{+})$ if $\gamma >1$ or $u_{0}(x)\in L_{1}^{2}(%
\mathbb{R}
^{+})\cap L_{-1}^{1}(%
\mathbb{R}
^{+})$ if $0<\gamma \leq 1$, with $\int_{0}^{\infty }xu_{0}(x)dx=0$ and $%
p(s)=2$. Then, the solution $u(x,t)$ to (\ref{eqn}) satisfies
\begin{eqnarray*}
\left\Vert u(x,t)\right\Vert _{L_{1}^{2}(%
\mathbb{R}
^{+})} &\leq &e^{-t}\left\Vert u_{0}(x)\right\Vert _{L_{1}^{2}(%
\mathbb{R}
^{+})},\ \ \gamma >2, \\
\left\Vert u(x,t)\right\Vert _{L_{1}^{2}(%
\mathbb{R}
^{+})} &\leq &Ce^{-t}\left\Vert u_{0}(x)\right\Vert _{L_{1}^{2}(%
\mathbb{R}
^{+})},\ \ 1<\gamma \leq 2, \\
\left\Vert u(x,t)\right\Vert _{L_{1}^{2}(%
\mathbb{R}
^{+})} &\leq &Ce^{-\gamma t}\left\Vert u_{0}(x)\right\Vert _{L_{1}^{2}(%
\mathbb{R}
^{+})},\ \ 0<\gamma \leq 1.
\end{eqnarray*}%
Let $p(s)$ a polynomial of degree $N$ (with $a_{0}>0$) and $\nu =\frac{p(1)-2%
}{\nu }$. If $\nu <0$ and $u_{0}(x)\in H_{1}^{\left\vert \nu \right\vert }(%
\mathbb{R}
^{+})$ then the solution $u(x,t)$ to (\ref{eqn}) satisfies
\begin{eqnarray*}
\left\Vert u(x,t)\right\Vert _{H_{1}^{\left\vert \nu \right\vert }(%
\mathbb{R}
^{+})} &\leq &Ce^{-t}\left\Vert u_{0}(x)\right\Vert _{H_{1}^{\left\vert \nu
\right\vert }(%
\mathbb{R}
^{+})},\ \ \gamma >1, \\
\left\Vert u(x,t)\right\Vert _{H_{1}^{\left\vert \nu \right\vert }(%
\mathbb{R}
^{+})} &\leq &Ce^{-\gamma t}\left\Vert u_{0}(x)\right\Vert
_{H_{1}^{\left\vert \nu \right\vert }(%
\mathbb{R}
^{+})},\ \ 0<\gamma \leq 1.
\end{eqnarray*}%
If $\nu \geq 0$ and $u_{0}(x)\in H_{1}^{\nu -\left[ \nu \right] }(%
\mathbb{R}
^{+})$ then the solution $u(x,t)$ to (\ref{eqn}) satisfies%
\begin{eqnarray*}
\left\Vert u(x,t)\right\Vert _{H_{1}^{\nu -\left[ \nu \right] }(%
\mathbb{R}
^{+})} &\leq &Ce^{-t}\left\Vert u_{0}(x)\right\Vert _{H_{1}^{\nu -\left[ \nu %
\right] }(%
\mathbb{R}
^{+})},\ \ \gamma >1, \\
\left\Vert u(x,t)\right\Vert _{H_{1}^{\nu -\left[ \nu \right] }(%
\mathbb{R}
^{+})} &\leq &Ce^{-\gamma t}\left\Vert u_{0}(x)\right\Vert _{H_{1}^{\nu -%
\left[ \nu \right] }(%
\mathbb{R}
^{+})},\ \ 0<\gamma \leq 1.
\end{eqnarray*}
\end{theorem}

\begin{figure}[tbp]
\centering\includegraphics[width=1.4\hsize]{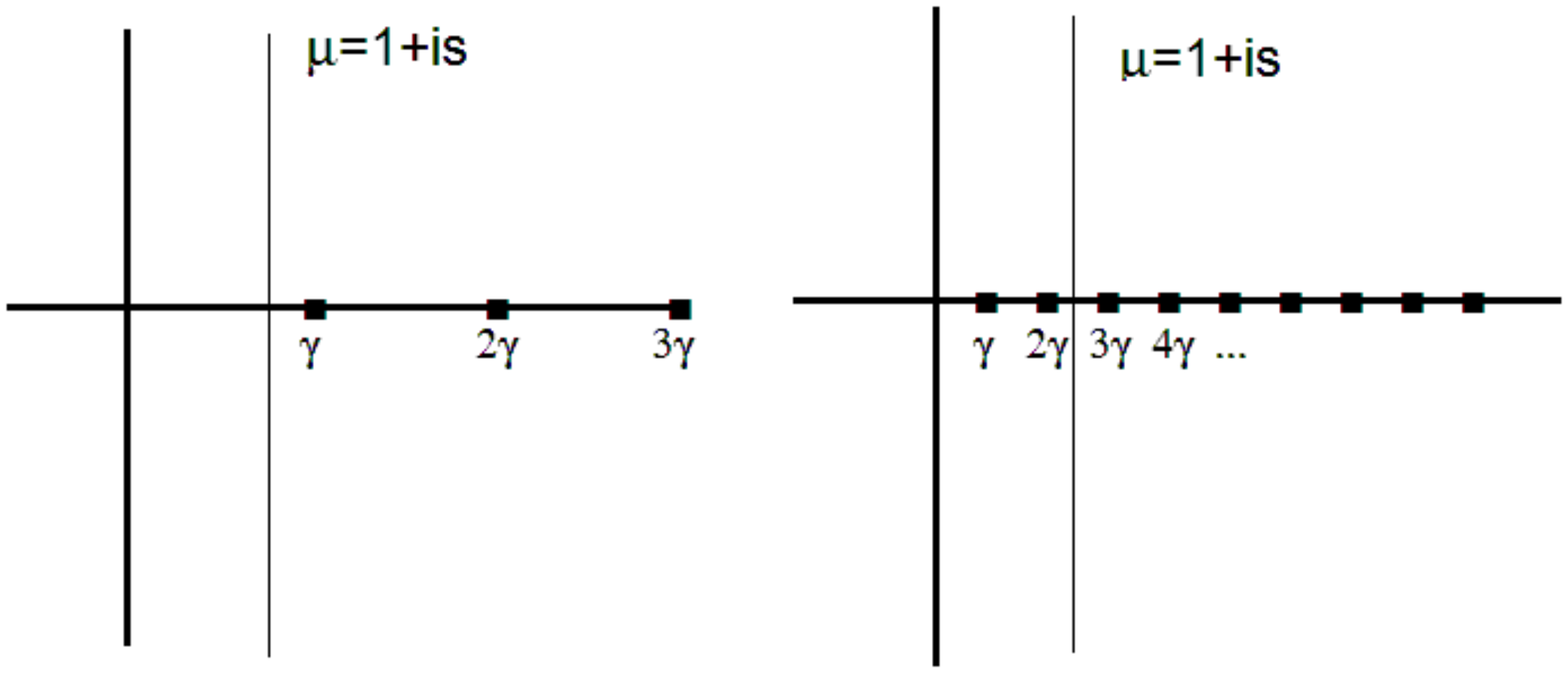}
\caption{ Spectrum of the operator $L$ in $L_{1}^{2}$. Note that the
discrete spectrum crosses the continuous spectrum at $\protect\gamma =1$.}
\label{fig}
\end{figure}

Central to the proof of these theorems is the description of the spectrum of
the linear operator $L$ defined as%
\begin{equation*}
Lu=\frac{\partial }{\partial x}(xu)+u-\gamma \left( \int_{x}^{\infty
}2y^{\gamma -1}u(y)dy-x^{\gamma }u(x)\right) ,
\end{equation*}%
corresponding to $p(s)=2$. The spectrum contains a discrete part $\left\{
n\gamma \right\} _{n=1}^{\infty }$ and, when the eigenvalue problem is
defined in $L_{1}^{2}(%
\mathbb{R}
^{+})$, a continuous part $1+is$, $s\in
\mathbb{R}
$ (see Figure \ref{fig}). The results will rely upon suitable sums over the
discrete spectrum and integrals along the continuous spectrum. Note the
presence of elements of the discrete spectrum in the interval $\left(
0,1\right) $ when $\gamma <1$. This is the origin of the different decay
rates in the last theorem.

Theorem \ref{th1} will be proved in section 3, while section 4 will be
devoted to the proof of theorems \ref{th2}, \ref{th3} and section 5 to the
proof of Theorem \ref{th4}.%
\begin{equation*}
\end{equation*}

\section{\protect\bigskip Selfsimilar solutions}

Selfsimilar solutions are functions independent of $t$ and therefore they
solve the stationary version of (\ref{eqnMellin}), i.e. the equation%
\begin{equation}
-\gamma z\widetilde{u}_{S}(z)+2\widetilde{u}_{S}(z)=\gamma \left( \frac{1}{%
\gamma }P(z)\widetilde{u}_{S}(z+1)-\widetilde{u}_{S}(z+1)\right) .
\label{ss}
\end{equation}%
The following property of Gamma functions%
\begin{equation*}
\Gamma (z+a+1)=(z+a)\Gamma (z+a),
\end{equation*}%
and the factorization of $K(z)=\gamma \left( \frac{1}{\gamma }P(z)-1\right) $
given by (\ref{fact}) allows to automatically check that
\begin{equation}
\widetilde{u}_{s}(z)=\frac{\prod_{i=1}^{N}\Gamma (z+\frac{i}{\gamma })}{%
\prod_{i=1}^{N}\Gamma (z-z_{i})}\Gamma (z),  \label{phiss}
\end{equation}%
is a solution to (\ref{ss}). In fact, following Theorem 2.2 in \cite{PM},
one can show that it is the only (up to a real \ and positive multiplicative
constant related to the total mass $\widetilde{u}_{S}(2)$) finite order
meromorphic function without divisors (zeros or poles) in $\Re (z)>0$ which
is a solution to equation (\ref{ss}). \ Formula (\ref{phiss}) is well-known
since the works of Treat \cite{Treat97a}, \cite{Treat97b} \ who realized it
represents the Mellin transform of a Meijer G-function and studied its
properties in detail. Here we present an alternative analysis that will make
more transparent our treatment of stability in later sections.

The function
\begin{equation}
F(z)=\frac{\prod_{i=1}^{N}\Gamma (z+\frac{i}{\gamma })}{\prod_{i=1}^{N}%
\Gamma (z-z_{i})},  \label{Fz}
\end{equation}%
has poles in the set $\left\{ -i/\gamma \right\} _{i=1}^{\infty }$ and is
analytic at $\Re (z)\geq 0$. It is simple to show that $\left\vert
F(z)\right\vert \leq C\left\vert z\right\vert ^{\nu }$ with $\nu
=\sum_{i}(i/\gamma +z_{i})$ for $\left\vert z\right\vert $ sufficiently
large. Hence, since $\left\vert F(z)\right\vert $ is analytic in $\Re
(z)\geq 0$, the inverse Mellin transform $f(s)$ is compactly supported in $%
\left[ 0,1\right] $. Take, for instance, $p(s)=a_{0}+a_{1}s$. Then%
\begin{equation*}
\frac{1}{\gamma }\sum_{i=0}^{1}\frac{a_{i}}{z+i/\gamma }-1=-\frac{(z-\frac{2%
}{\gamma })(z+\frac{1}{2\gamma }a_{0})}{z\left( z+\frac{1}{\gamma }\right) },
\end{equation*}%
where we have used $a_{0}\frac{1}{2}+a_{1}\frac{1}{3}=1$. We obtain then%
\begin{equation}
\widetilde{u}_{S}(z)=\frac{\Gamma (z+\frac{1}{\gamma })}{\Gamma (z+\frac{1}{%
2\gamma }a_{0})}\Gamma (z).  \label{dmt}
\end{equation}%
When $a_{0}>2$ (and hence $p(s)$ is a decreasing function), we can compute
the inverse Mellin transform of%
\begin{equation*}
F(z)=\frac{\Gamma (z+\frac{1}{\gamma })}{\Gamma (z+\frac{1}{2\gamma }a_{0})},
\end{equation*}%
using (\ref{p4}):%
\begin{equation*}
f(s)=\frac{1}{\Gamma (\frac{a_{0}-2}{2\gamma })}s^{\frac{1}{\gamma }}(1-s)^{%
\frac{a_{0}-2}{2\gamma }-1},
\end{equation*}%
the inverse Mellin transform of
\begin{equation*}
G(z)=\Gamma (z)
\end{equation*}%
using (\ref{p1}), and then use the convolution formula (\ref{p0}) to
conclude with the following expression for the selfsimilar solution:%
\begin{eqnarray*}
u_{S}(x) &=&\frac{1}{\Gamma (\frac{a_{0}-2}{2\gamma })}\int_{1}^{\infty }s^{-%
\frac{1}{\gamma }}(1-s^{-1})^{\frac{a_{0}-2}{2\gamma }-1}s^{-1}e^{-sx}ds \\
&=&\frac{1}{\Gamma (\frac{a_{0}-2}{2\gamma })}\int_{1}^{\infty }s^{-\frac{%
a_{0}}{2\gamma }}(s-1)^{\frac{a_{0}-2}{2\gamma }-1}e^{-sx}ds.
\end{eqnarray*}%
We can compute the asymptotic behavior at infinity of $u_{S}(x)$ by noting

\begin{eqnarray}
u_{S}(x) &=&\frac{e^{-x}}{\Gamma (\frac{a_{0}-2}{2\gamma })}\int_{1}^{\infty
}s^{-\frac{a_{0}}{2\gamma }}(s-1)^{\frac{a_{0}-2}{2\gamma }-1}e^{-(s-1)x}ds
\notag \\
&=&\frac{e^{-x}}{\Gamma (\frac{a_{0}-2}{2\gamma })}\int_{0}^{\infty
}(1+\sigma )^{-\frac{a_{0}}{2\gamma }}\sigma ^{\frac{a_{0}-2}{2\gamma }%
-1}e^{-\sigma x}d\sigma  \notag \\
&=&\frac{e^{-x}x^{-\frac{a_{0}-2}{2\gamma }}}{\Gamma (\frac{a_{0}-2}{2\gamma
})}\int_{0}^{\infty }(1+\frac{\sigma }{x})^{-\frac{a_{0}}{2\gamma }}\sigma ^{%
\frac{a_{0}-2}{2\gamma }-1}e^{-\sigma }d\sigma  \notag \\
&=&e^{-x}x^{-\frac{a_{0}-2}{2\gamma }}(1+O(x^{-1}))\ ,\ \ \text{as }%
x\rightarrow \infty .  \label{h}
\end{eqnarray}%
On the other hand,%
\begin{eqnarray*}
u_{S}(0) &=&\frac{1}{\Gamma (\frac{a_{0}-2}{2\gamma })}\int_{1}^{\infty }s^{-%
\frac{1}{\gamma }}(1-s^{-1})^{\frac{a_{0}-2}{2\gamma }-1}s^{-1}ds \\
&=&\frac{1}{\Gamma (\frac{a_{0}-2}{2\gamma })}\int_{0}^{1}\sigma ^{\frac{1}{%
\gamma }-1}(\sigma -1)^{\frac{a_{0}-2}{2\gamma }-1}d\sigma =\frac{\Gamma (%
\frac{1}{\gamma })}{\Gamma (\frac{a_{0}}{2\gamma })}.
\end{eqnarray*}%
If $a_{0}<2$, we can always write $\frac{1}{\gamma }+\nu =k+\frac{1}{2\gamma
}a_{0}$ with $k$ a positive integer and $\nu \in \left( 0,1\right) $.
Therefore, we can write%
\begin{eqnarray*}
&&\frac{\Gamma (z+k+\frac{1}{2\gamma }a_{0}-\nu )}{\Gamma (z+\frac{1}{%
2\gamma }a_{0})}\Gamma (z) \\
&=&\frac{\Gamma (z+\frac{1}{2\gamma }a_{0}-\nu )}{\Gamma (z+\frac{1}{2\gamma
}a_{0})}\left( z+\frac{1}{2\gamma }a_{0}-\nu \right) ...\left( z+(k-1)+\frac{%
1}{2\gamma }a_{0}-\nu \right) \Gamma (z),
\end{eqnarray*}%
and since the inverse Mellin transform of%
\begin{equation*}
G(z)=\left( z+\frac{1}{2\gamma }a_{0}-\nu \right) ...\left( z+(k-1)+\frac{1}{%
2\gamma }a_{0}-\nu \right) \Gamma (z),
\end{equation*}%
is%
\begin{equation*}
g(x)=(-1)^{k}x^{\frac{1}{2\gamma }a_{0}-\nu +k}\frac{d^{k}}{dx^{k}}\left(
x^{-\frac{1}{2\gamma }a_{0}+\nu }e^{-x}\right) \sim x^{k}e^{-x},\ \text{as }%
x\rightarrow \infty ,
\end{equation*}%
and the inverse Mellin transform of
\begin{equation*}
F(z)=\frac{\Gamma (z+\frac{1}{2\gamma }a_{0}-\nu )}{\Gamma (z+\frac{1}{%
2\gamma }a_{0})},
\end{equation*}%
is%
\begin{equation*}
f(s)=\frac{1}{\Gamma (\nu )}s^{\frac{1}{2\gamma }a_{0}-\nu }(1-s)^{\nu -1},
\end{equation*}%
for $s<1$ and zero for $s\geq 1$, we can repeat the asymptotic analysis in (%
\ref{h}) to obtain%
\begin{equation}
u_{S}(x)\sim e^{-x}x^{\frac{2-a_{0}}{2\gamma }}(1+O(x^{-1}))\ ,\ \ \text{as }%
x\rightarrow \infty .  \label{admt}
\end{equation}%
Note that we can apply the same analysis leading from (\ref{dmt}) to the
asymptotics for its inverse Mellin transform (\ref{admt}) in case $a_{0}$
were a complex number and separating the cases $\Re (a_{0})<2$ and $\Re
(a_{0})>2$ in identical form.

The analysis above can also be extended to polynomials $p(s)$ of degree
greater than one by writing%
\begin{equation*}
F(z)=\prod_{i=1}^{N}F_{i}(z),
\end{equation*}%
with%
\begin{equation*}
F_{i}(z)=\frac{\Gamma (z+\frac{i}{\gamma })}{\Gamma (z-z_{i})},
\end{equation*}%
and defining%
\begin{eqnarray*}
\widetilde{u}_{S}^{1}(z) &=&F_{1}(z)\Gamma (z) \\
\widetilde{u}_{S}^{i}(z) &=&F_{i}(z)\widetilde{u}_{s}^{i-1}(z),\ i=2,...,N-1
\\
\widetilde{u}_{S}(z) &=&F_{N}(z)\widetilde{u}_{s}^{N-1}(z),
\end{eqnarray*}%
so that
\begin{eqnarray*}
u_{S}^{1}(x) &\sim &e^{-x}x^{\frac{1}{\gamma }+z_{1}}\ ,\ \ \text{as }%
x\rightarrow \infty  \\
&&.... \\
u_{S}(x) &\sim &e^{-x}x^{\sum_{i=1}^{N}\left( \frac{i}{\gamma }+z_{i}\right)
}=e^{-x}x^{\nu },\ \ \text{as }x\rightarrow \infty ,
\end{eqnarray*}%
where we have estimated at each step as in (\ref{h}) We can finally use (\ref%
{nu}) to show%
\begin{equation*}
u_{S}(x)\sim x^{\frac{p(1)-2}{\gamma }}e^{-x}(1+O(x^{-1})),\ \ \text{as }%
x\rightarrow \infty .
\end{equation*}

Note now that, by (\ref{estf1}) in the Appendix B,
\begin{equation}
\left\vert \prod_{i=1}^{N}F_{i}\left( i\lambda +\frac{1}{2}\right)
\right\vert \leq C(1+\lambda ^{2})^{\frac{p(1)-2}{2}},  \label{ff}
\end{equation}%
and hence%
\begin{equation*}
\left\vert \widetilde{u}_{S}\left( i\lambda +\frac{1}{2}\right) \right\vert
\leq C(1+\lambda ^{2})^{\frac{p(1)-2}{2}}\frac{\sqrt{\pi }}{\sqrt{\cosh
\lambda }},
\end{equation*}%
where we have used%
\begin{equation*}
\left\vert \Gamma \left( i\lambda +\frac{1}{2}\right) \right\vert
^{2}=\Gamma \left( i\lambda +\frac{1}{2}\right) \Gamma \left( -i\lambda +%
\frac{1}{2}\right) =\frac{\pi }{\cosh \lambda },
\end{equation*}%
so that, for any $n\in
\mathbb{N}
$,
\begin{equation*}
\left\Vert \widetilde{u}_{S}\right\Vert _{N}^{2}\equiv \int_{-\infty
}^{\infty }(1+\lambda ^{2})^{N}\left\vert \widetilde{u}_{S}\left( i\lambda +%
\frac{1}{2}\right) \right\vert ^{2}\leq C_{N}.
\end{equation*}%
Now we use the equivalence in norms%
\begin{equation*}
\left\Vert \widetilde{u}_{S}\right\Vert _{N}^{2}\sim \int_{0}^{\infty
}\left( \left\vert u_{s}(x)\right\vert ^{2}+x^{2N}\left\vert \frac{d^{N}u_{s}%
}{dx^{N}}(x)\right\vert ^{2}\right) ,
\end{equation*}%
and the fact that%
\begin{equation*}
\sup_{x}\left\vert x^{i+\frac{1}{2}}\frac{d^{i}u_{S}}{dx^{i}}(x)\right\vert
^{2}\leq C\int_{0}^{\infty }\left( \left\vert u_{S}(x)\right\vert
^{2}+x^{2N}\left\vert \frac{d^{N}u_{S}}{dx^{N}}(x)\right\vert ^{2}\right) ,\
i=0,1,..,N-1,
\end{equation*}%
to conclude that $u_{S}(x)\in C^{\infty }(0,\infty )$ and hence completing
the proof of Theorem \ref{th1}. Notice that, according to the inverse Mellin
transform formula (\ref{invm}), $u_{S}(0)$ is $\lim_{z\rightarrow 0}z%
\widetilde{u}_{S}(z)$, i.e. the residual of $\widetilde{u}_{S}(z)$ at $z=0$.
That is,%
\begin{equation*}
u_{S}(0)=\frac{\prod_{i=1}^{N}\Gamma (\frac{i}{\gamma })}{%
\prod_{i=1}^{N}\Gamma (-z_{i})}.
\end{equation*}%
The fact that the other poles of $\widetilde{u}_{S}(z)$ are $\left\{
-j-i/\gamma \right\} $ with $i=0,1,..,N$ and $j=0,1,...,$ implies that $%
u_{S}(x)$ is expanded in terms of the form $x^{j+i/\gamma }$ so that
derivatives will not be bounded at the origin in general.

\section{Stability of selfsimilar solutions}

In this section we will study the stability of selfsimilar solutions, that
is the decay (or not) in time of $u^{\ast }(x,t)=u(x,t)-u_{S}(x)$. As
metioned in the introduction, $u^{\ast }(x,t)$ will satisfy (\ref{eqn}) and,
in addition, the mass conservation condition:%
\begin{equation*}
\int_{0}^{\infty }xu^{\ast }(x,t)dx=0.
\end{equation*}%
In particular, the initial data $u^{\ast }(x,0)$ must have zero mass. In
what follows, we will drop for notational simplicity the $\ast $, and will
study equation (\ref{eqn}) for zero mass initial data or, equivalently,
equation (\ref{eqnt}) with initial data $u_{0}(x)$ such that%
\begin{equation*}
\int_{0}^{\infty }x^{\frac{2}{\gamma }-1}u_{0}(x)dx=0.
\end{equation*}

In order to study the stability of selfsimilar solutions, we have to
consider the evolution problem (\ref{eqnMellin}) and look for solutions in
the form%
\begin{equation*}
\widetilde{u}(z,t)=e^{-\mu t}\widetilde{U}(z),
\end{equation*}%
to yield the eigenvalue problem%
\begin{equation*}
-\mu \widetilde{U}(z)-\gamma z\widetilde{U}(z)+2\widetilde{U}(z)=\gamma
\left( \frac{1}{\gamma }P(z)-1\right) \widetilde{U}(z).
\end{equation*}

\subsection{Proof of Theorem \protect\ref{th2}}

We consider first the case $p(s)=2$ and the eigenvalue problem%
\begin{equation}
\left( -\mu -\gamma z+2\right) \widetilde{U}(z)=\left( \frac{2}{z}-\gamma
\right) \widetilde{U}(z+1),  \label{eigenvalue}
\end{equation}%
with solution%
\begin{equation}
\widetilde{U}(z)=\frac{\Gamma \left( -z+\frac{2}{\gamma }+1\right) }{\Gamma
\left( -z+\frac{2-\mu }{\gamma }+1\right) }\Gamma \left( z\right) .
\label{fi1}
\end{equation}%
The inverse Mellin transform (formula \ref{p3}) yields then%
\begin{equation*}
U(x)=\Gamma \left( \frac{2}{\gamma }+1\right) M\left( \frac{2}{\gamma }+1,%
\frac{2-\mu }{\gamma }+1,-x\right) ,
\end{equation*}%
and using (\ref{asy}) we find%
\begin{equation*}
M\left( \frac{2}{\gamma }+1,\frac{2-\mu }{\gamma }+1,-x\right) \sim \Gamma
\left( \frac{2-\mu }{\gamma }+1\right) \left( \frac{e^{-x}(-x)^{\frac{\mu }{%
\gamma }}}{\Gamma (\frac{2-\mu }{\gamma }+1)}+\frac{x^{-\left( \frac{2}{%
\gamma }+1\right) }}{\Gamma (-\frac{\mu }{\gamma })}\right) ,\text{ as }%
x\rightarrow \infty .
\end{equation*}%
In order for the solutions to decay exponentially fast at infinity we must
impose%
\begin{equation*}
\mu =\mu _{n}=n\gamma ,\ n\in
\mathbb{N}
,
\end{equation*}%
and then%
\begin{eqnarray*}
\widetilde{U}(z) &=&\widetilde{U}_{n}(z)=\frac{\Gamma \left( -z+\frac{2}{%
\gamma }+1\right) }{\Gamma \left( -z+\frac{2}{\gamma }-n+1\right) }\Gamma
\left( z\right) \\
&=&\left( -z+\frac{2}{\gamma }\right) \left( -z+\frac{2}{\gamma }-1\right)
...\left( -z+\frac{2}{\gamma }-(n-1)\right) \Gamma \left( z\right) \\
&=&(-1)^{n}\left( z-\frac{2}{\gamma }\right) \left( z-\frac{2}{\gamma }%
+1\right) ...\left( z-\frac{2}{\gamma }+(n-1)\right) \Gamma \left( z\right) .
\end{eqnarray*}%
One can easily verify that the inverse Mellin transform of $\widetilde{U}%
_{n}(z)$ (see \ref{p51}, \ref{p52}) is
\begin{equation*}
U_{n}(x)=x^{n-\frac{2}{\gamma }}\frac{d^{n}}{dx^{n}}\left( x^{\frac{2}{%
\gamma }}e^{-x}\right) ,
\end{equation*}%
by simple integration by parts:%
\begin{eqnarray*}
&&\int_{0}^{\infty }x^{z-1}x^{n-\frac{2}{\gamma }}\frac{d^{n}}{dx^{n}}\left(
x^{\frac{2}{\gamma }}e^{-x}\right) dx \\
&=&-\left( z-\frac{2}{\gamma }+(n-1)\right) \int_{0}^{\infty }x^{z-1}x^{n-1-%
\frac{2}{\gamma }}\frac{d^{n-1}}{dx^{n-1}}\left( x^{\frac{2}{\gamma }%
}e^{-x}\right) dx \\
&=&...=(-1)^{n}\left( z-\frac{2}{\gamma }\right) \left( z-\frac{2}{\gamma }%
+1\right) ...\left( z-\frac{2}{\gamma }+(n-1)\right) \Gamma \left( z\right) .
\end{eqnarray*}%
By Rodrigues formula for generalized Laguerre polynomials we have%
\begin{equation}
U_{n}(x)=n!L_{n}^{\left( \frac{2}{\gamma }-n\right) }(x)e^{-x}.  \label{fi2}
\end{equation}%
It is worth mentioning that the particular solution to the evolution problem
$u(x,t)=e^{-\gamma t}U_{1}(x)$ was found before in the particular case when $%
\gamma =\frac{2}{1+n}$ by rather different methods involving infinite power
series expansions (see example 2.3.3 in \cite{B1}).

We remark that the eigenvalue equation (\ref{eigenvalue}) with eigenvalue $%
\mu _{n}=n\gamma $ possesses a second solution
\begin{equation*}
\widetilde{U}_{n}(z)=\frac{\Gamma \left( z-\frac{2}{\gamma }+n\right) \Gamma
\left( -z+\frac{2}{\gamma }+1\right) }{\Gamma \left( 1-z\right) },
\end{equation*}%
that leads, after direct calculation to the eigenfunction%
\begin{equation*}
W_{n}(x)=\frac{1}{\Gamma (1-\frac{2}{\gamma }+\left[ \frac{2}{\gamma }\right]
)}x^{n-\frac{2}{\gamma }}\frac{d^{n}}{dx^{n}}\left( x^{\left[ \frac{2}{%
\gamma }\right] +1}\int_{0}^{1}e^{-xu}(1-u)^{-\frac{2}{\gamma }+\left[ \frac{%
2}{\gamma }\right] }du\right) ,
\end{equation*}%
which present, for general $\gamma $, power law decay $O(x^{\left[ \frac{2}{%
\gamma }\right] -\frac{2}{\gamma }})$ which has unbounded first moment.

We cannot expect that the set $\left\{ U_{n}(x)\right\} _{n=1}^{\infty }$
forms an orthogonal set (in some functional space), since the
transport-fragmentation operator $L$ is not self-adjoint. Nevertheless, we
will see that expansions in $\left\{ U_{n}(x)\right\} _{n=1}^{\infty }$ are
possible and the set is complete in a suitable functional space. We start by
assuming $\gamma \in \left\{ 2,2/2,...2/i,...\right\} $\ so that certain
algebraic relations to be used hold true. Notice first that due to the
recurrence relation (formula 22.12.6 evaluated at $y=0$ combined with 22.4.7
in \cite{A})%
\begin{equation}
L_{j}^{(\frac{2}{\gamma }-m)}(x)=\sum_{i=0}^{m}\left(
\begin{array}{c}
n-m+j-i-1 \\
j-i%
\end{array}%
\right) L_{i}^{(\frac{2}{\gamma }-n)}(x),  \label{aba}
\end{equation}%
where $n>m$, and the orthogonality relation (22.2.12 in \cite{A}):
\begin{equation*}
\int_{0}^{\infty }x^{\alpha }L_{n}^{(\alpha )}(x)L_{m}^{(\alpha
)}(x)e^{-x}dx=\frac{\Gamma (n+\alpha +1)}{n!}\delta _{n,m}\ ,
\end{equation*}%
we have%
\begin{eqnarray*}
\int_{0}^{\infty }x^{\frac{2}{\gamma }-n}L_{n}^{(\frac{2}{\gamma }%
-n)}(x)L_{m}^{(\frac{2}{\gamma }-m)}(x)e^{-x}dx &=&0,n>m \\
\int_{0}^{\infty }x^{\frac{2}{\gamma }-n}L_{n}^{(\frac{2}{\gamma }%
-n)}(x)L_{n}^{(\frac{2}{\gamma }-n)}(x)e^{-x}dx &=&\frac{\Gamma (\frac{2}{%
\gamma }+1)}{n!} \\
\int_{0}^{\infty }x^{\frac{2}{\gamma }-1}(L_{1}^{\frac{2}{\gamma }%
-1}(x))^{2}e^{-x}dx &=&\Gamma (\frac{2}{\gamma }+1),
\end{eqnarray*}%
where we have used the restriction $\gamma =2/i$ so that the integrals
converge. It also holds%
\begin{equation}
L_{n}^{(\frac{2}{\gamma })}(x)=\sum_{i=0}^{n}\left(
\begin{array}{c}
n \\
n-i%
\end{array}%
\right) L_{i}^{(\frac{2}{\gamma }-i)}(x),  \label{ft1}
\end{equation}%
which can be easily verified by expanding both sides of the equation in
terms of $L_{j}(x)$ using (\ref{aba}) with $n=\frac{2}{\gamma }$ and then
noting that the coefficients of each Laguerre polynomial at both sides of
the equaition are identical due to the Chu--Vandermonde identity for
binomial coefficients.

We formally write, using (\ref{ft1}),%
\begin{eqnarray*}
u_{0}(x) &=&\sum_{j=1}^{\infty }\alpha _{j}L_{j}^{\left( \frac{2}{\gamma }%
\right) }(x)e^{-x}=\sum_{j=0}^{\infty }\alpha _{j}\left( \sum_{i=0}^{j}\frac{%
j!}{i!(j-i)!}L_{i}^{(\frac{2}{\gamma }-i)}(x)e^{-x}\right) \\
&=&\sum_{i=0}^{\infty }\left( \sum_{j=i}^{\infty }\alpha _{j}\frac{j!}{%
i!(j-i)!}\right) L_{i}^{(\frac{2}{\gamma }-i)}(x)e^{-x},
\end{eqnarray*}%
with%
\begin{equation*}
\alpha _{n}=\frac{1}{\Gamma (\frac{2}{\gamma }+1)}\int_{0}^{\infty }x^{\frac{%
2}{\gamma }}L_{n}^{(\frac{2}{\gamma })}(x)u_{0}(x)dx,
\end{equation*}%
so that we can also write%
\begin{equation*}
u_{0}(x)=\sum_{i=0}^{\infty }a_{i}U_{i}(x)=\sum_{i=0}^{\infty
}a_{i}i!L_{i}^{\left( \frac{2}{\gamma }-i\right) }(x)e^{-x}
\end{equation*}%
with
\begin{equation*}
a_{i}i!=\sum_{j=i}^{\infty }\alpha _{j}\frac{j!}{i!(j-i)!}.
\end{equation*}%
Next we will deduce conditions on the initial data $u_{0}(x)$ under which
the formal series
\begin{equation*}
u(x,t)=\sum_{i=0}^{\infty }a_{i}e^{-i\gamma t}U_{i}(x)
\end{equation*}%
converges absolutely away from the origin. First we note

\begin{eqnarray}
u(x,t) &=&\sum_{i=0}^{\infty }a_{i}e^{-i\gamma t}(-1)^{\frac{2}{\gamma }%
-n}x^{i-\frac{2}{\gamma }}\Gamma \left( \frac{2}{\gamma }+1\right)
L_{2/\gamma }^{(i-\frac{2}{\gamma })}(x)e^{-x}  \notag \\
&=&\sum_{i=0}^{\infty }a_{i}e^{-i\gamma t}(-1)^{\frac{2}{\gamma }-i}\frac{%
d^{2/\gamma }}{dx^{2/\gamma }}\left( x^{i}e^{-x}\right)   \label{tocho1}
\end{eqnarray}%
where we have used the relation%
\begin{equation*}
x^{\frac{2}{\gamma }-n}L_{n}^{(\frac{2}{\gamma }-n)}(x)=\frac{\Gamma
(2/\gamma +1)}{n!}(-1)^{\frac{2}{\gamma }-n}L_{2/\gamma }^{(n-\frac{2}{%
\gamma })}(x),
\end{equation*}%
as well as Rodrigues formula. Secondly,
\begin{equation}
\left\vert a_{i}\right\vert \leq \frac{1}{(i!)^{2}}\left( \sum_{j=i}^{\infty
}\left\vert \alpha _{j}\right\vert j^{i}\right) \leq \frac{1}{(i!)^{2}}%
\left( \sum_{j=i}^{\infty }\frac{1}{j^{2}}\right) ^{\frac{1}{2}}\left(
\sum_{j=i}^{\infty }j^{2i+2}\left\vert \alpha _{j}\right\vert ^{2}\right) ^{%
\frac{1}{2}}\leq \frac{C}{(i!)^{2}}\left\Vert u_{0}\right\Vert _{\mathbb{H}%
^{i+1}}  \label{tocho2}
\end{equation}%
where we have denoted
\begin{equation*}
\left\Vert u_{0}\right\Vert _{\mathbb{H}^{k}}^{2}=\sum_{j=0}^{\infty
}(1+j)^{2k}\left\vert \alpha _{j}\right\vert ^{2}.
\end{equation*}%
Hence, from (\ref{tocho1}) and (\ref{tocho2}),%
\begin{eqnarray}
\left\vert u(x,t)\right\vert  &\leq &\sum_{i=0}^{\infty }\left\vert
a_{i}\right\vert \left\vert \frac{d^{2/\gamma }}{dx^{2/\gamma }}\left(
x^{i}e^{-x}\right) \right\vert e^{-i\gamma t}  \notag \\
&\leq &C\sum_{i=0}^{\infty }\frac{1}{(i!)^{2}}\left\Vert u_{0}\right\Vert _{%
\mathbb{H}^{i+1}}(i+1)^{\frac{2}{\gamma }}(1+x^{i})e^{-x}e^{-i\gamma t}
\label{tocho3}
\end{eqnarray}%
and since
\begin{equation*}
\sum_{i=0}^{\infty }\frac{1}{(i!)^{1+\sigma }}(1+x^{i})\leq Ce^{\alpha x},
\end{equation*}%
for $\alpha $ arbitrarily small and a suitable $C$ depending on $\alpha $,
we can guarantee
\begin{equation}
\left\vert u(x,t)\right\vert \leq Ce^{-\gamma t}e^{-(1-\alpha )x},
\label{estimat}
\end{equation}%
provided%
\begin{equation*}
\left\Vert u_{0}\right\Vert _{\mathbb{H}^{i}}\leq C(i!)^{1+\sigma }.
\end{equation*}%
By using the identity $xL_{n}^{\left( \frac{2}{\gamma }\right) \prime
}(x)=nL_{n}^{(\frac{2}{\gamma })}(x)-(n+\frac{2}{\gamma })L_{n-1}^{(\frac{2}{%
\gamma })}(x)$ we find after integration by parts%
\begin{eqnarray*}
\int_{0}^{\infty }x^{\frac{2}{\gamma }}L_{n}^{(\frac{2}{\gamma })}(x)x\frac{%
du_{0}(x)}{dx}dx &=&-\left( \frac{2}{\gamma }+1+n\right) \int_{0}^{\infty
}x^{\frac{2}{\gamma }}L_{n}^{(\frac{2}{\gamma })}(x)u_{0}(x)dx \\
&&+\left( n+\frac{2}{\gamma }\right) \int_{0}^{\infty }x^{\frac{2}{\gamma }%
}L_{n-1}^{\left( \frac{2}{\gamma }\right) }(x)u_{0}(x)dx
\end{eqnarray*}%
and hence%
\begin{equation}
\int_{0}^{\infty }x^{\frac{2}{\gamma }}\left\vert \left( x\frac{d}{dx}%
\right) ^{k}u_{0}(x)\right\vert ^{2}e^{x}dx\leq C\left\Vert u_{0}\right\Vert
_{\mathbb{H}^{k}}^{2}  \label{gev}
\end{equation}%
which charaterizes $u_{0}$ as belonging to the $(1+\sigma )$ Sobolev-Gevrey
class.

The solution to (\ref{eqnt}) with $p(s)=2$ is then%
\begin{eqnarray*}
u(x,t) &=&\sum_{n=1}^{\infty }\left( \frac{1}{\Gamma \left( \frac{2}{\gamma }%
+1\right) }\int_{0}^{\infty }x^{\frac{2}{\gamma }-n}L_{n}^{(\frac{2}{\gamma }%
-n)}(x)u_{0}(x)dx\right) a_{n}e^{-n\gamma t}n!L_{n}^{\left( \frac{2}{\gamma }%
-n\right) }(x)e^{-x}, \\
a_{n} &=&\sum_{j=n}^{\infty }\frac{j!}{(n!)^{2}(j-n)!}\frac{1}{\Gamma (\left[
\frac{2}{\gamma }\right] +1)}\int_{0}^{\infty }y^{\frac{2}{\gamma }%
}L_{j}^{\left( \frac{2}{\gamma }\right) }(y)u_{0}(y)dy.
\end{eqnarray*}%
If we lift now the restriction $\gamma =2/k$ and write
\begin{equation*}
\frac{2}{\gamma }=k+\delta ,\ k\in
\mathbb{N}
,\ \delta \in (0,1),
\end{equation*}%
we can then find%
\begin{equation*}
\widetilde{U}_{n}(z)=\frac{(-1)^{n}\Gamma \left( z\right) }{\Gamma \left(
z+\delta \right) }\left[ \left( z-\frac{2}{\gamma }\right) \left( z-\frac{2}{%
\gamma }+1\right) ...\left( z-\frac{2}{\gamma }+(n-1)\right) \Gamma \left( z-%
\frac{2}{\gamma }+k\right) \right] ,
\end{equation*}%
so that the inverse Mellin transform of
\begin{equation*}
\Psi _{n}(z)=\frac{\Gamma \left( z-\delta \right) }{\Gamma \left( z\right) }%
\widetilde{U}_{n}(z),
\end{equation*}%
is%
\begin{equation*}
V_{n}(x)=x^{1-\frac{2}{\gamma }+(n-1)}\frac{d^{n}}{dx^{n}}\left(
x^{k}e^{-x}\right) =n!x^{-\delta }L_{n}^{(k-n)}(x)e^{-x}.
\end{equation*}%
Hence, using (\ref{p4}), we conclude
\begin{equation*}
L_{n}^{(k-n)}(x)e^{-x}=\frac{1}{\Gamma (\delta )}\left( \int_{x}^{\infty
}(1-x/y)^{\delta -1}y^{\delta -1}L_{n}^{(k+\delta -n)}(y)e^{-y}dy\right) .
\end{equation*}%
Then, defining%
\begin{equation*}
v_{0}(x)=\frac{1}{x^{\delta }\Gamma (\delta )}\left( \int_{x}^{\infty
}(1-x/y)^{\delta -1}y^{\delta -1}u_{0}(y)dy\right) ,
\end{equation*}%
we have%
\begin{equation*}
v_{0}(x)=x^{-\delta }\sum_{n=1}^{\infty }a_{n}n!L_{n}^{(k-n)}(x)e^{-x},
\end{equation*}%
so that%
\begin{equation*}
a_{n}=\sum_{j=i}^{\infty }\frac{j!}{(n!)^{2}(j-n)!}\frac{1}{\Gamma (k+1)}%
\int_{0}^{\infty }x^{\frac{2}{\gamma }}L_{j}^{(k)}(x)v_{0}(x)dx,
\end{equation*}%
For $m=0,1,2,...$ we have%
\begin{eqnarray}
\int x^{m}x^{\frac{2}{\gamma }}(v_{0}(x))^{2}dx &=&\int_{-\infty }^{\infty
}\left\vert \widetilde{v_{0}}(i\lambda +\frac{m+1}{2}+k+\delta )\right\vert
^{2}d\lambda  \notag \\
&=&\int_{-\infty }^{\infty }\left\vert \frac{\Gamma \left( i\lambda +\frac{%
m+1}{2}+k\right) }{\Gamma \left( i\lambda +\frac{m+1}{2}+k+\delta \right) }%
\right\vert ^{2}\left\vert \widetilde{u_{0}}(i\lambda +\frac{m}{2}+k+\delta
)\right\vert ^{2}d\lambda  \notag \\
&\leq &C\int x^{m}x^{\frac{2}{\gamma }}(u_{0}(x))^{2}dx,  \label{xm}
\end{eqnarray}%
where we have bounded, since $\nu \in (0,1)$, $\left\vert \frac{\Gamma
\left( i\lambda +\frac{m+1}{2}+k\right) }{\Gamma \left( i\lambda +\frac{m+1}{%
2}+k+\delta \right) }\right\vert ^{2}\leq C$ uniformly in $\lambda $ and $m$
(see (\ref{estf1}) in the Appendix B). Hence%
\begin{eqnarray*}
\int e^{x}x^{\frac{2}{\gamma }}(v_{0}(x))^{2}dx &=&\sum_{m=0}^{\infty }\frac{%
1}{m!}\int x^{m}x^{\frac{2}{\gamma }}(v_{0}(x))^{2}dx \\
&\leq &C\sum_{m=0}^{\infty }\frac{1}{m!}\int x^{m}x^{\frac{2}{\gamma }%
}(u_{0}(x))^{2}dx=C\int e^{x}x^{\frac{2}{\gamma }}(u_{0}(x))^{2}dx.
\end{eqnarray*}%
The argument can be reproduced for any $(x\frac{d}{dx})^{j}v_{0}(x)$ to
conclude (\ref{estimat}) provided $u_{0}(x)$ is in the $(1+\sigma )$
Sobolev-Gevrey class.

We conclude then the formula%
\begin{equation}
u(x,t)=\sum_{n=1}^{\infty }a_{n}e^{-n\gamma t}n!L_{n}^{\left( \left[ \frac{2%
}{\gamma }\right] -n\right) }(x)e^{-x}.  \label{expresion}
\end{equation}%
with%
\begin{equation*}
a_{n}=\sum_{j=n}^{\infty }\frac{j!}{(n!)^{2}(j-n)!}\frac{1}{\Gamma (\left[
\frac{2}{\gamma }\right] +1)}\int_{0}^{\infty }x^{\frac{2}{\gamma }%
}L_{j}^{\left( \left[ \frac{2}{\gamma }\right] \right) }(x)v_{0}(x)dx.
\end{equation*}

The expression (\ref{expresion}) allows to estimate $u(x,t)$ in various
norms and describe its asymptotic behaviors explicitly. First, we can
estimate $u(x,t)$ as in (\ref{estimat}). We can obtain similar estimates for
the derivatives of $u(x,t)$ by using (\ref{tocho1}):%
\begin{equation*}
\frac{\partial ^{r}u}{\partial x^{r}}(x,t)=\sum_{i=0}^{\infty
}a_{i}e^{-i\gamma t}(-1)^{\frac{2}{\gamma }-i}\frac{d^{\left[ 2/\gamma %
\right] +r}}{dx^{\left[ 2/\gamma \right] +r}}\left( x^{i}e^{-x}\right)
\end{equation*}%
and estimating as in (\ref{tocho3}):%
\begin{eqnarray*}
\left\vert \frac{\partial ^{r}u}{\partial x^{r}}(x,t)\right\vert &\leq
&\sum_{i=0}^{\infty }\left\vert a_{i}\right\vert \left\vert \frac{d^{\left[
2/\gamma \right] +r}}{dx^{\left[ 2/\gamma \right] +r}}\left(
x^{i}e^{-x}\right) \right\vert e^{-i\gamma t} \\
&\leq &C\sum_{i=0}^{\infty }\frac{1}{(i!)^{2}}\left\Vert u_{0}\right\Vert _{%
\mathbb{H}^{i+1}}(i+1)^{\frac{2}{\gamma }+r}(1+x^{i})e^{-x}e^{-i\gamma t}
\end{eqnarray*}%
so that%
\begin{equation}
\left\vert \frac{\partial ^{r}u}{\partial x^{r}}(x,t)\right\vert \leq
C_{r}e^{-\gamma t}e^{-(1-\alpha )x},  \label{estu}
\end{equation}%
for $\alpha >0$ and a suitable constant $C$. This completes the proof of
Theorem \ref{th2}.

\subsection{Proof of Theorem \protect\ref{th3}}

Finally, in order to prove Theorem \ref{th3}, we consider a general
polynomial $p(s)$ and we note that estimate (\ref{xm}) also holds when
\begin{equation*}
\widetilde{v_{0}}(z)=F(z)\widetilde{u_{0}}(z),
\end{equation*}%
provided $F(z)$ is analytic in $\Re z>0$ and $\left\vert F(i\lambda +\frac{m%
}{2})\right\vert \leq C$ uniformly in $\lambda $ and $m$. This also implies,
since%
\begin{equation}
\widetilde{\left( \left( x\frac{d}{dx}\right) ^{k}v_{0}\right) }(z)=F(z)%
\widetilde{\left( \left( x\frac{d}{dx}\right) ^{k}u_{0}\right) }(z)
\label{mellinxdx}
\end{equation}%
so that%
\begin{equation}
\int e^{x}x^{\frac{2}{\gamma }}\left( \left( x\frac{d}{dx}\right)
^{k}v_{0}(x)\right) ^{2}dx\leq C\int e^{x}x^{\frac{2}{\gamma }}\left( \left(
x\frac{d}{dx}\right) ^{k}u_{0}(x)^{2}\right) dx,  \label{exp}
\end{equation}%
and hence, $v_{0}(x)$ is in a $(1+\sigma )$ Sobolev-Gevrey class if $%
u_{0}(x) $ is in the same class.

In particular, if%
\begin{equation*}
F(z)=\frac{\prod_{i=1}^{N}\Gamma (z-z_{i})}{\prod_{i=1}^{N}\Gamma (z+\frac{i%
}{\gamma })},
\end{equation*}%
and the regularity index $\nu =\sum_{i}^{N}\left( \frac{i}{\gamma }%
+z_{i}\right) >0,$ we have (see \ref{estfff})%
\begin{equation}
\left\vert F(i\lambda +\frac{m}{2}+\frac{2}{\gamma })\right\vert \leq
C(1+\lambda ^{2})^{-\frac{1}{2}\sum_{i}^{N}\left( \frac{i}{\gamma }%
+z_{i}\right) }=C(1+\lambda ^{2})^{-\frac{p(1)-2}{2\gamma }}\leq C.
\label{estm}
\end{equation}%
When $p(1)\geq 2$, then estimate (\ref{exp}) holds true.

In the case when the regularity index $\nu $ is negative (i.e. $p(1)<2$), we
cannot rely on (\ref{estm}), but on using (\ref{estff}) instead. Then%
\begin{eqnarray*}
\int e^{x}x^{\frac{2}{\gamma }}(v_{0}(x))^{2}dx &=&\sum_{m=0}^{\infty }\frac{%
1}{m!}\int x^{m}x^{\frac{2}{\gamma }}(v_{0}(x))^{2}dx \\
&=&\sum_{m=0}^{\infty }\frac{1}{m!}\int_{-\infty }^{\infty }\left\vert
\widetilde{v_{0}}\left( i\lambda +\frac{m+1}{2}+\frac{2}{\gamma }\right)
\right\vert ^{2}d\lambda \\
&=&\sum_{m=0}^{\infty }\frac{1}{m!}\int_{-\infty }^{\infty }\left\vert
F\left( i\lambda +\frac{m}{2}+\frac{2}{\gamma }\right) \right\vert
^{2}\left\vert \widetilde{u_{0}}\left( i\lambda +\frac{m+1}{2}+\frac{2}{%
\gamma }\right) \right\vert ^{2}d\lambda \\
&\leq &C\sum_{m=0}^{\infty }\frac{1}{m!}\int_{-\infty }^{\infty
}(1+m)^{2\left\vert \nu \right\vert }(1+\lambda ^{2})^{\left\vert \nu
\right\vert }\left\vert \widetilde{u_{0}}\left( i\lambda +\frac{m+1}{2}+%
\frac{2}{\gamma }\right) \right\vert ^{2}d\lambda \\
&\leq &C\sum_{m=0}^{\infty }\frac{1}{m!}\int_{-\infty }^{\infty
}(1+m^{2}+\lambda ^{2})^{\left[ \left\vert \nu \right\vert \right]
+1}\left\vert \widetilde{u_{0}}\left( i\lambda +\frac{m+1}{2}+\frac{2}{%
\gamma }\right) \right\vert ^{2}d\lambda \\
&\leq &C\int e^{x}x^{\frac{2}{\gamma }}\left[ (u_{0}(x))^{2}+\left( \left( x%
\frac{d}{dx}\right) ^{\left[ \left\vert \nu \right\vert \right]
+1}u_{0}(x)\right) ^{2}\right] dx,
\end{eqnarray*}%
and using (\ref{mellinxdx}) we deduce in general%
\begin{eqnarray}
&&\int e^{x}x^{\frac{2}{\gamma }}\left( \left( x\frac{d}{dx}\right)
^{k}v_{0}(x)\right) ^{2}dx  \notag \\
&\leq &C\int e^{x}x^{\frac{2}{\gamma }}\left[ \left( \left( x\frac{d}{dx}%
\right) ^{k}u_{0}(x)\right) ^{2}+(\left( x\frac{d}{dx}\right) ^{k+\left[
\left\vert \nu \right\vert \right] +1}u_{0}(x))^{2}\right] dx.  \label{exp2}
\end{eqnarray}%
Hence, by (\ref{gev}), and if $u_{0}(x)$ is in a Sobolev-Gevrey class $%
1+\sigma $ then
\begin{eqnarray*}
\int_{0}^{\infty }x^{\frac{2}{\gamma }}\left\vert (x\frac{d}{dx})^{k+\left[
\left\vert \nu \right\vert \right] +1}u_{0}(x)\right\vert ^{2}e^{x}dx &\leq
&C\left\Vert u_{0}\right\Vert _{\mathbb{H}^{k+\left[ \left\vert \nu
\right\vert \right] +1}}^{2}\leq C((k+\left[ \left\vert \nu \right\vert %
\right] +1)!)^{2(1+\sigma )} \\
&\leq &C(k!)^{2(1+\sigma ^{\prime })},
\end{eqnarray*}%
with $0<\sigma <\sigma ^{\prime }$ and $v_{0}(x)$ is in a Sobolev-Gevrey
class $1+\sigma ^{\prime }$. Now we realize that\bigskip
\begin{equation*}
\widetilde{v}(z,t)=F(z)\widetilde{u}(z,t),
\end{equation*}%
where $v(x,t)$ is provided by (\ref{expresion})$:$%
\begin{equation*}
v(x,t)=\sum_{n=1}^{\infty }a_{n}e^{-n\gamma t}n!L_{n}^{\left( \frac{2}{%
\gamma }-n\right) }(x)e^{-x},
\end{equation*}%
with%
\begin{equation*}
a_{n}=\sum_{j=n}^{\infty }\frac{j!}{(n!)^{2}(j-n)!}\frac{1}{\Gamma (\left[
\frac{2}{\gamma }\right] +1)}\int_{0}^{\infty }x^{\frac{2}{\gamma }%
}L_{j}^{\left( \left[ \frac{2}{\gamma }\right] \right) }(x)w_{0}(x)dx,
\end{equation*}%
and%
\begin{equation*}
w_{0}(x)=\frac{1}{x^{\delta }\Gamma (\nu )}\left( \int_{x}^{\infty
}(1-x/y)^{\delta -1}y^{\delta -1}v_{0}(y)dy\right) ,
\end{equation*}%
so that the previous Theorem applies and, in particular, inequality (\ref%
{estu}) holds. If $\nu \leq 0$ and $s>0$ then%
\begin{eqnarray*}
\left\vert \widetilde{u}(i\lambda +s,t)\right\vert &\leq &\frac{1}{%
\left\vert F(i\lambda +s)\right\vert }\left\vert \widetilde{v}(i\lambda
+s,t)\right\vert \\
&\leq &C(1+\lambda ^{2})^{\frac{p(1)-2}{2\gamma }}\left\vert \widetilde{v}%
(i\lambda +s,t)\right\vert \leq C\left\vert \widetilde{v}(i\lambda
+s,t)\right\vert ,
\end{eqnarray*}%
which yields, by Plancherel's formula,%
\begin{equation*}
\int_{0}^{\infty }x^{2s-1}\left\vert u(x,t)\right\vert ^{2}dx\leq
C_{s}\int_{0}^{\infty }x^{2s-1}\left\vert v(x,t)\right\vert ^{2}dx,
\end{equation*}%
and, analogously for any $r>1$, $s>0$%
\begin{equation*}
\int_{0}^{\infty }x^{2r+2s-1}\left\vert \frac{\partial ^{r}u}{\partial x^{r}}%
(x,t)\right\vert ^{2}dx\leq C_{r,s}\int_{0}^{\infty }x^{2r+2s-1}\left\vert
\frac{\partial ^{r}v}{\partial x^{r}}(x,t)\right\vert ^{2}dx.
\end{equation*}%
Therefore%
\begin{eqnarray*}
&&\int_{0}^{\infty }x^{2s+2r-1}\left\vert \frac{\partial ^{r}}{\partial x^{r}%
}\left( e^{\frac{x}{2}}u(x,t)\right) \right\vert ^{2}dx \\
&\leq &C\sum_{i\leq j\leq r}\int_{0}^{\infty }e^{x}x^{2s+2j-1}\left\vert
\frac{\partial ^{i}}{\partial x^{i}}u(x,t)\right\vert ^{2}dx \\
&\leq &C\sum_{i\leq j\leq r}\int_{0}^{\infty }e^{x}x^{2s+2j-1}\left\vert
\frac{\partial ^{i}}{\partial x^{i}}v(x,t)\right\vert ^{2}dx\leq
Ce^{-2\gamma t},
\end{eqnarray*}%
where we have used (\ref{estu}) to estimate the norm of $v(x,t)$. We note
now
\begin{equation*}
\int_{0}^{\infty }x^{2s+2r-1}\left\vert \frac{\partial ^{r}f(x)}{\partial
x^{r}}\right\vert ^{2}dx=\int\limits_{-\infty }^{\infty }(1+\left\vert
\lambda \right\vert ^{2r})\left\vert \int\limits_{0}^{\infty }x^{i\lambda
+s-1}f(x)dx\right\vert ^{2}d\lambda
\end{equation*}%
and standard Sobolev embeddings $H^{r}\subset L^{\infty }$, $r\geq 1$ then
yield%
\begin{equation}
x^{s}\left\vert u(x,t)\right\vert \leq Ce^{-\gamma t}e^{-\frac{1}{2}x}.
\label{uf}
\end{equation}%
The argument can be extended to any $x\frac{d}{dx}$-derivative of $u(x,t)$
and estimate (\ref{uf}) will hold for a suitable constant $C$ if $u$ is
replaced by $x^{r}\frac{\partial ^{r}u}{\partial x^{r}}$ so that $u\in
C^{\infty }(0,\infty )$.

If $\nu >0$, then taking $r\geq 1$,%
\begin{eqnarray*}
&&\sup_{x}\left\vert x^{s}e^{\frac{x}{2}}u(x,t)\right\vert ^{2} \\
&\leq &\int_{0}^{\infty }x^{2s+2r-1}\left\vert \frac{\partial ^{r}}{\partial
x^{r}}\left( e^{\frac{x}{2}}u(x,t)\right) \right\vert ^{2}dx \\
&\leq &C\sum_{i\leq j\leq r}\int_{0}^{\infty }e^{x}x^{2s+2j-1}\left\vert
\frac{\partial ^{i}}{\partial x^{i}}u(x,t)\right\vert ^{2}dx \\
&\leq &C\sum_{i\leq j\leq r}\int_{0}^{\infty }e^{x}x^{2s+2j-1}(1+x^{2(\left[
\nu \right] +1)})\left\vert \frac{\partial ^{i+\left[ \nu \right] +1}}{%
\partial x^{i+\left[ \nu \right] +1}}v(x,t)\right\vert ^{2}dx\leq
Ce^{-2\gamma t},
\end{eqnarray*}%
where we have applied (\ref{estu}), so that estimate (\ref{uf}) also holds
in this case. Analogously, for $r\geq k+1$%
\begin{eqnarray*}
&&\sup_{x}\left\vert x^{s}e^{\frac{x}{2}}x^{k}\frac{\partial ^{k}u}{\partial
x^{k}}(x,t)\right\vert ^{2} \\
&\leq &C\sum_{i\leq j\leq r}\int_{0}^{\infty }e^{x}x^{2s+2j+2k-1}\left\vert
\frac{\partial ^{i}}{\partial x^{i}}u(x,t)\right\vert ^{2}dx \\
&\leq &C\sum_{i\leq j\leq r}\int_{0}^{\infty }e^{x}x^{2s+2j+2k-1}(1+x^{2(
\left[ \nu \right] +1)})\left\vert \frac{\partial ^{i+k+\left[ \nu \right]
+1}}{\partial x^{i+k+\left[ \nu \right] +1}}v(x,t)\right\vert ^{2}dx \\
&\leq &C_{r}e^{-2\gamma t},
\end{eqnarray*}%
where we have applied (\ref{estu}), so that $u\in C^{\infty }(0,\infty )$.
This concludes the proof of Theorem \ref{th3}.

\section{The evolution problem in $L_{1}^{2}(%
\mathbb{R}
^{+})$}

In this section we will study the fragmentation problem for initial data
with power-like decay at infinity and prove Theorem \ref{th4}. Following
previous works, we will assume an initial data $u(x)$ with zero mass and
such that $u(x)\in L_{1}^{2}(%
\mathbb{R}
^{+})$, that is square integrable functions with the measure $d\mu =xdx$. In
the variable $x^{\gamma }$, the measure is $d\mu =x^{\frac{2}{\gamma }-1}dx$
so we will take, for the fragmentation problem in this new variable, $%
u(x)\in L_{\frac{2}{\gamma }-1}^{2}(%
\mathbb{R}
^{+})$. As mentioned in the previous section, when $p(s)=2$, a solution to
the eigenvalue problem%
\begin{equation*}
Lu\equiv \gamma x\frac{\partial u}{\partial x}+2u-\gamma \left( \frac{2}{%
\gamma }\int_{x}^{\infty }u(y)pdy-xu(x)\right) =\mu u,
\end{equation*}%
in the Mellin transform formulation\ is
\begin{equation*}
\widetilde{U}_{\mu }(z)=\frac{\Gamma \left( z-\frac{2-\mu }{\gamma }\right)
\Gamma \left( -z+\frac{2}{\gamma }+1\right) }{\Gamma \left( 1-z\right) },
\end{equation*}%
and since%
\begin{equation*}
\int_{0}^{\infty }x^{z-1}x^{c}M(a,b,-x)dx=\frac{\Gamma \left( z+c\right)
\left( a-z-c\right) }{\Gamma \left( b-z-c\right) }
\end{equation*}%
by choosing%
\begin{equation*}
a=\frac{2}{\gamma }+\frac{1}{\gamma }\left( \mu -2\right) +1,b=\frac{1}{%
\gamma }\left( \mu -2\right) +1,c=\frac{1}{\gamma }\left( \mu -2\right) ,
\end{equation*}%
we have in $x-$space%
\begin{equation*}
u_{\mu }(x)=x^{\frac{1}{\gamma }\left( \mu -2\right) }M\left( \frac{2}{%
\gamma }+\frac{1}{\gamma }\left( \mu -2\right) +1,\frac{1}{\gamma }\left(
\mu -2\right) +1,-x\right) .
\end{equation*}%
Note the following asymptotics
\begin{eqnarray*}
&&x^{\frac{1}{\gamma }\left( \mu -2\right) }M\left( \frac{2}{\gamma }+\frac{1%
}{\gamma }\left( \mu -2\right) +1,\frac{1}{\gamma }\left( \mu -2\right)
+1,-x\right) \\
&\sim &\left\{
\begin{array}{c}
B(\frac{2}{\gamma }+\frac{1}{\gamma }\left( \mu -2\right) +1,1-\frac{2}{%
\gamma })x^{\frac{1}{\gamma }\left( \mu -2\right) },\ \text{as }x\rightarrow
0 \\
\Gamma (\frac{2}{\gamma }+\frac{1}{\gamma }\left( \mu -2\right) +1)x^{-(1+%
\frac{2}{\gamma })},\ \text{as }x\rightarrow \infty%
\end{array}%
\right. ,
\end{eqnarray*}%
so that%
\begin{equation*}
u_{\mu }\in L_{\frac{2}{\gamma }+\frac{2\nu }{\gamma }-1}^{2}(%
\mathbb{R}
^{+})\text{ if }\mu >1-\nu .
\end{equation*}%
We take now%
\begin{equation*}
\mu =\mu _{0}=1+is,
\end{equation*}%
with $s\in
\mathbb{R}
$ and note that

\begin{equation*}
Lu_{\mu }-\mu _{0}u_{\mu }=Lu_{\mu }-\mu u_{\mu }+(\mu -\mu _{0})u_{\mu },
\end{equation*}%
with%
\begin{equation*}
\mu =\mu _{0}-\varepsilon ,\ \varepsilon >0.
\end{equation*}%
Then%
\begin{eqnarray*}
\left\Vert Lu_{\mu }-\mu _{0}u_{\mu }\right\Vert _{L_{\frac{2}{\gamma }%
-1}^{2}}^{2} &=&\varepsilon ^{2}\left\Vert u_{\mu }\right\Vert _{L_{\frac{2}{%
\gamma }-1}^{2}}^{2} \\
&\leq &C\varepsilon ^{2}\left( \int_{0}^{1}x^{\frac{2}{\gamma }-1}\left\vert
x^{\frac{1}{\gamma }\left( 1+\varepsilon +is-2\right) }\right\vert
^{2}dx+1\right) \\
&\leq &C\varepsilon ^{2}\left( \int_{0}^{1}x^{-1+\frac{2\varepsilon }{\gamma
}}dx+1\right) =O(\varepsilon )\rightarrow 0\text{ as }\varepsilon
\rightarrow 0^{+},
\end{eqnarray*}%
which characterizes $\mu _{0}$ as an element of the continuous spectrum of $%
L $ in $L_{\frac{2}{\gamma }-1}^{2}(%
\mathbb{R}
^{+})$. Analogously, one can show that $1-\nu +is$, $s\in
\mathbb{R}
$ is continuous spectrum in $L_{\frac{2}{\gamma }+\frac{2\nu }{\gamma }%
-1}^{2}(%
\mathbb{R}
^{+})$. The solution of the eigenvalue problem for $\mu =1+is$ and $\gamma
>2 $ is%
\begin{equation*}
u_{s}(x)=\frac{1}{\Gamma (1-\frac{2}{\gamma })}x^{\frac{1}{\gamma }\left(
-1+is\right) }\left( \int_{0}^{1}e^{-xw}w^{\frac{1}{\gamma }+\frac{1}{\gamma
}is}(1-w)^{-\frac{2}{\gamma }}dw\right) ,
\end{equation*}%
since

\begin{eqnarray*}
&&\int_{0}^{\infty }x^{z-1}x^{\frac{1}{\gamma }\left( -1+is\right) }\left(
\int_{0}^{1}e^{-xw}w^{\frac{1}{\gamma }+\frac{1}{\gamma }is}(1-w)^{-\frac{2}{%
\gamma }}dw\right) dx \\
&=&\int_{0}^{\infty }v^{z+\frac{1}{\gamma }\left( -1+is\right)
-1}e^{-v}\left( \int_{0}^{1}w^{\frac{1}{\gamma }+\frac{1}{\gamma }is}w^{-z-%
\frac{1}{\gamma }\left( -1+is\right) }(1-w)^{-\frac{2}{\gamma }}dw\right) dv
\\
&=&\Gamma (1-\frac{2}{\gamma })\frac{\Gamma (z+\frac{1}{\gamma }\left(
-1+is\right) )\Gamma (-z+\frac{2}{\gamma }+1)}{\Gamma (1-z)}.
\end{eqnarray*}%
If we write now the following superposition (in the form of integral in $s$)%
\begin{equation*}
u(x)=\int_{-\infty }^{\infty }u_{s}(x)A(s)ds,
\end{equation*}%
we formally find that defining%
\begin{eqnarray*}
G(xw) &=&\int_{-\infty }^{\infty }(xw)^{\frac{1}{\gamma }is}A(s)ds, \\
F(v) &=&e^{-v}G(v),
\end{eqnarray*}%
one can write%
\begin{equation*}
u(x)=\frac{1}{\Gamma (1-\frac{2}{\gamma })}x^{-\frac{1}{\gamma }}\left(
\int_{0}^{1}F(xw)w^{\frac{1}{\gamma }}(1-w)^{-\frac{2}{\gamma }}dw\right) ,
\end{equation*}%
with the following relation between $u$ and $F$:%
\begin{eqnarray}
\int_{0}^{\infty }x^{z+\frac{1}{\gamma }-1}u(x)dx &=&\frac{1}{\Gamma (1-%
\frac{2}{\gamma })}\int_{0}^{\infty }x^{z-1}\left( \int_{0}^{1}F(xw)w^{\frac{%
1}{\gamma }}(1-w)^{-\frac{2}{\gamma }}dw\right) dx  \notag \\
&=&\frac{1}{\Gamma (1-\frac{2}{\gamma })}\int_{0}^{\infty }v^{z-1}F(v)\left(
\int_{0}^{1}w^{-z}w^{\frac{1}{\gamma }}(1-w)^{-\frac{2}{\gamma }}dw\right) dv
\\
&=&\frac{\Gamma (-z+1+\frac{1}{\gamma })}{\Gamma (-z+2-\frac{1}{\gamma })}%
\int_{0}^{\infty }v^{z-1}F(v)dv,  \label{id}
\end{eqnarray}%
so that one can write%
\begin{equation*}
u(x)=x^{-1}I_{1-\frac{2}{\gamma }}(x^{\frac{1}{\gamma }}F(x)),
\end{equation*}%
where $I_{\alpha }$ is the $\alpha $ Riemann-Liouville fractional integral
defined by%
\begin{equation*}
I_{\alpha }g(x)=\frac{1}{\Gamma (\alpha )}\int_{0}^{x}\frac{g(y)dy}{%
(x-y)^{1-\alpha }}.
\end{equation*}%
The solution to the evolution problem is then%
\begin{equation*}
u(x,t)=\int_{-\infty }^{\infty }u_{s}(x)e^{-(1+is)t}A(s)ds,
\end{equation*}%
and then%
\begin{equation*}
G(xwe^{-\gamma t})=\int_{-\infty }^{\infty }(xwe^{-\gamma t})^{\frac{1}{%
\gamma }is}A(s)ds,
\end{equation*}%
so that we can write%
\begin{equation}
u(x,t)=\frac{1}{\Gamma (1-\frac{2}{\gamma })}e^{-t}x^{-\frac{1}{\gamma }%
}\int_{0}^{1}e^{-xw(1-e^{-\gamma t})}F(e^{-\gamma t}xw)w^{\frac{1}{\gamma }%
}(1-w)^{-\frac{2}{\gamma }}dw,  \label{fxt}
\end{equation}%
providing an explicit expression for the solution to the fragmentation
problem. In the Appendix C we will alternatively show that (\ref{fxt}) is a
solution by direct replacement into the equation. One can now split $u(x,0)$
in its positive and negative parts $u^{\pm }(x,0)$ and $F(s)$ into the
corresponding $F^{\pm }(s)$. We can therefore estimate
\begin{eqnarray}
&&\int_{0}^{\infty }x^{\frac{2}{\gamma }-1}\left\vert u(x,t)\right\vert
^{2}dx  \notag \\
&=&\int_{0}^{\infty }x^{\frac{2}{\gamma }-1}\left\vert \frac{1}{\Gamma (1-%
\frac{2}{\gamma })}e^{-t}x^{-\frac{1}{\gamma }}\int_{0}^{1}e^{-xw(1-e^{-%
\gamma t})}F(e^{-\gamma t}xw)w^{\frac{1}{\gamma }}(1-w)^{-\frac{2}{\gamma }%
}dw\right\vert ^{2}dx  \notag \\
&=&e^{-2t}\int_{0}^{\infty }s^{\frac{2}{\gamma }-1}\left\vert \frac{1}{%
\Gamma (1-\frac{2}{\gamma })}s^{-\frac{1}{\gamma }}\int_{0}^{1}e^{-sw(e^{%
\gamma t}-1)}F(sw)w^{\frac{1}{\gamma }}(1-w)^{-\frac{2}{\gamma }%
}dw\right\vert ^{2}ds  \notag \\
&\leq &e^{-2t}\int_{0}^{\infty }s^{\frac{2}{\gamma }-1}\left\vert \frac{1}{%
\Gamma (1-\frac{2}{\gamma })}s^{-\frac{1}{\gamma }}\int_{0}^{1}\left\vert
F(sw)\right\vert w^{\frac{1}{\gamma }}(1-w)^{-\frac{2}{\gamma }%
}dw\right\vert ^{2}ds  \notag \\
&\leq &e^{-2t}\int_{0}^{\infty }x^{\frac{2}{\gamma }-1}\left\vert
u(x,0)\right\vert ^{2}dx,  \label{l21}
\end{eqnarray}%
where we have used the norm inequality%
\begin{eqnarray*}
\left\Vert F\right\Vert &\equiv &\left[ \int_{0}^{\infty }s^{\frac{2}{\gamma
}-1}\left\vert \frac{1}{\Gamma (1-\frac{2}{\gamma })}s^{-\frac{1}{\gamma }%
}\int_{0}^{1}\left\vert F(sw)\right\vert w^{\frac{1}{\gamma }}(1-w)^{-\frac{2%
}{\gamma }}dw\right\vert ^{2}ds\right] ^{\frac{1}{2}} \\
&\leq &\left\Vert F^{+}\right\Vert +\left\Vert F^{-}\right\Vert =\left[
\int_{0}^{\infty }x^{\frac{2}{\gamma }-1}\left\vert u^{+}(x,0)\right\vert
^{2}dx\right] ^{\frac{1}{2}}+\left[ \int_{0}^{\infty }x^{\frac{2}{\gamma }%
-1}\left\vert u^{-}(x,0)\right\vert ^{2}dx\right] ^{\frac{1}{2}}.
\end{eqnarray*}%
Analogously, one can prove%
\begin{equation*}
\int_{0}^{\infty }x^{\frac{2+2\nu }{\gamma }-1}\left\vert u(x,t)\right\vert
^{2}dx\leq e^{-2(1-\nu )t}\int_{0}^{\infty }x^{\frac{2+2\nu }{\gamma }%
-1}\left\vert u(x,0)\right\vert ^{2}dx,
\end{equation*}%
for $\nu \in \left( -1,1\right) $. Estimates similar to (\ref{l21}) also
hold for derivatives of $u$. By applying $\left( x\frac{d}{dx}\right) ^{n}$
to (\ref{fxt}) and noting%
\begin{eqnarray*}
\left( x\frac{d}{dx}\right) ^{n}\left( x^{-\frac{1}{\gamma }%
}e^{-xw(1-e^{-\gamma t})}F(e^{-\gamma t}xw)\right) &=&x^{-\frac{1}{\gamma }%
}H(e^{-\gamma t}xw,t) \\
H(v,t) &=&\sum_{i\leq n}a_{i}(v,t)v^{i}F^{(i)}(v),
\end{eqnarray*}%
with $\left\vert a_{i}(v,t)\right\vert $ bounded, together with the identity
(analogous to (\ref{id})):%
\begin{eqnarray*}
&&\int_{0}^{\infty }x^{z+\frac{1}{\gamma }+i-1}u^{(i)}(x)dx \\
&=&\frac{(z+\frac{1}{\gamma }+i-1)(z+\frac{1}{\gamma }+i-2)\ldots (z+\frac{1%
}{\gamma })}{(z+i-1)(z+i-2)\ldots z}\frac{\Gamma (-z+1+\frac{1}{\gamma })}{%
\Gamma (-z+2-\frac{1}{\gamma })}\int_{0}^{\infty }v^{z+i-1}F^{(i)}(v)dv,
\end{eqnarray*}%
so that
\begin{equation*}
\left\vert \int_{0}^{\infty }v^{z+i-1}F^{(i)}(v)dv\right\vert \leq
C\left\vert \frac{\Gamma (-z+2-\frac{1}{\gamma })}{\Gamma (-z+1+\frac{1}{%
\gamma })}\right\vert \left\vert \int_{0}^{\infty }x^{z+\frac{1}{\gamma }%
+i-1}u^{(i)}(x)dx\right\vert ,
\end{equation*}%
and then%
\begin{equation}
\int_{0}^{\infty }x^{\frac{2}{\gamma }+2i-1}\left\vert \frac{\partial ^{i}}{%
\partial x^{i}}u(x,t)\right\vert ^{2}dx\leq
Ce^{-2t}\sum_{j=0}^{i}\int_{0}^{\infty }x^{\frac{2}{\gamma }+2j-1}\left\vert
\frac{\partial ^{j}}{\partial x^{j}}u(x,0)\right\vert ^{2}dx.  \label{estd}
\end{equation}

When $\gamma \in \left( 1,2\right] $, the solutions to the eigenvalue
problem (in Mellin transform representation) can be written as%
\begin{equation*}
\widetilde{U}_{\mu }(z)=-\left( z-\frac{2}{\gamma }\right) \frac{\Gamma
\left( z-\frac{1}{\gamma }+\frac{is}{\gamma }\right) \Gamma \left( -z+\frac{2%
}{\gamma }\right) }{\Gamma \left( 1-z\right) },
\end{equation*}%
and we can expand%
\begin{equation*}
u^{\ast }(x,0)=\frac{1}{\Gamma (2-\frac{2}{\gamma })}x^{-\frac{1}{\gamma }%
}\left( \int_{0}^{1}F(xw)w^{\frac{1}{\gamma }}(1-w)^{1-\frac{2}{\gamma }%
}dw\right) ,
\end{equation*}%
where%
\begin{equation*}
\widetilde{u^{\ast }}(z,t)=-\frac{\widetilde{u}(z,t)}{\left( z-\frac{2}{%
\gamma }\right) }.
\end{equation*}%
Proceeding as above, we can prove%
\begin{equation*}
\int_{0}^{\infty }x^{\frac{2}{\gamma }-1}\left\vert u^{\ast
}(x,t)\right\vert ^{2}dx\leq e^{-2t}\int_{0}^{\infty }x^{\frac{2}{\gamma }%
-1}\left\vert u^{\ast }(x,0)\right\vert ^{2}dx,
\end{equation*}%
and estimating derivatives as in (\ref{estd}) one can obtain%
\begin{equation*}
\int_{0}^{\infty }x^{\frac{2}{\gamma }-1}\left\vert u(x,t)\right\vert
^{2}dx\leq Ce^{-2t}\int_{0}^{\infty }x^{\frac{2}{\gamma }-1}\left\vert
u(x,0)\right\vert ^{2}dx.
\end{equation*}%
Finally, we look at $\gamma \leq 1$. In this case, $\left[ \frac{1}{\gamma }%
\right] $ eigenvalues in the set $\left\{ n\gamma \right\} _{n=1}^{\infty }$
are smaller or equal than $1$ and their values are $\mu =\gamma $, $2\gamma $%
,...$\left[ \frac{1}{\gamma }\right] \gamma $. We will label the
corresponding eigenfunctions decaying exponentially fast at infinity as $%
u^{i}(x)$ with $\mu _{i}=\gamma i$. The solution to the eigenvalue problem is%
\begin{eqnarray*}
\widetilde{U}_{\mu }(z) &=&(-1)^{\left[ \frac{2}{\gamma }\right] }\left( z-%
\frac{2}{\gamma }\right) \left( z+1-\frac{2}{\gamma }\right) ...\left( z+%
\left[ \frac{2}{\gamma }\right] -1-\frac{2}{\gamma }\right) \\
&&\times \frac{\Gamma \left( z-\frac{1}{\gamma }+\frac{is}{\gamma }\right)
\Gamma \left( -z+\frac{2}{\gamma }-\left[ \frac{2}{\gamma }\right] +1\right)
}{\Gamma \left( 1-z\right) },
\end{eqnarray*}%
and we can write%
\begin{equation*}
u^{\ast }(x,0)=\frac{1}{\Gamma (1+\left[ \frac{2}{\gamma }\right] -\frac{2}{%
\gamma })}x^{-\frac{1}{\gamma }}\left( \int_{0}^{1}F(xw)w^{\frac{1}{\gamma }%
}(1-w)^{\left[ \frac{2}{\gamma }\right] -\frac{2}{\gamma }}dw\right) ,
\end{equation*}%
where%
\begin{equation*}
\widetilde{u^{\ast }}(z,0)=-\frac{\widetilde{u}(z,0)-\sum_{i=1}^{\left[
\frac{2}{\gamma }\right] -1}a_{i}\widetilde{u^{i}}(z)}{(-1)^{\left[ \frac{2}{%
\gamma }\right] }\left( z-\frac{2}{\gamma }\right) \left( z+1-\frac{2}{%
\gamma }\right) ...\left( z+\left[ \frac{2}{\gamma }\right] -1-\frac{2}{%
\gamma }\right) },
\end{equation*}%
with $\left\{ a_{i}\right\} $ suitably chosen so that $\widetilde{u^{\ast }}%
(z)$ has no poles at $z=\frac{2}{\gamma }$, $\frac{2}{\gamma }-1$,...,$\frac{%
2}{\gamma }-\left[ \frac{2}{\gamma }\right] +1$. This is achieved by solving
the system%
\begin{eqnarray*}
\widetilde{u}(\frac{2}{\gamma }-1,0)-\Gamma \left( \frac{2}{\gamma }%
-1\right) a_{1} &=&0 \\
\widetilde{u}(\frac{2}{\gamma }-2,0)-2\Gamma \left( \frac{2}{\gamma }%
-2\right) a_{1}+(2\cdot 1)\Gamma \left( \frac{2}{\gamma }-2\right) a_{2} &=&0
\\
\widetilde{u}(\frac{2}{\gamma }-3,0)-3\Gamma \left( \frac{2}{\gamma }%
-3\right) a_{1}+(3\cdot 2)\Gamma \left( \frac{2}{\gamma }-3\right)
a_{2}-3!\Gamma \left( \frac{2}{\gamma }-3\right) a_{3} &=&0 \\
&&...
\end{eqnarray*}%
where we are assuming the boundedness the moments%
\begin{equation*}
\widetilde{u}(\frac{2}{\gamma }-j,0)=\int_{0}^{\infty }x^{\frac{2}{\gamma }%
-j-1}u(x,0)dx,\ j=1,...,\left[ \frac{2}{\gamma }\right] -1
\end{equation*}%
a fact guaranteed if $u(x,0)\in L_{-1}^{1}(%
\mathbb{R}
^{+})\cap L_{\frac{2}{\gamma }-1}^{2}(%
\mathbb{R}
^{+})$.

Then, proceeding as above, we estimate%
\begin{equation*}
\int_{0}^{\infty }x^{\frac{2}{\gamma }-1}\left\vert u^{\ast
}(x,t)\right\vert ^{2}dx\leq Ce^{-2t}\int_{0}^{\infty }x^{\frac{2}{\gamma }%
-1}\left\vert u^{\ast }(x,0)\right\vert ^{2}dx.
\end{equation*}%
Nevertheless, the leading asymptotic behavior for $u(x,t)$ is provided by
the first eigenfunctions multiplied by $e^{-\mu _{1}t}=e^{-\gamma t}$ so that%
\begin{equation*}
\int_{0}^{\infty }x^{\frac{2}{\gamma }-1}\left\vert u(x,t)\right\vert
^{2}dx\leq Ce^{-2\gamma t}.
\end{equation*}

For a general polynomial daughter fragment distribution function $p(s)$,
similar estimates hold depending on the parameter $\nu =\frac{p(1)-2}{\nu }$%
. If we define the Mellin transform $\widetilde{g}(z,t)$ of a function $%
g(x,t)$ defined by
\begin{equation*}
\widetilde{g}(z,t)=F(z)\widetilde{u}(z,t),
\end{equation*}%
where $F(z)$ is defined in (\ref{Fz}), then $g(x,t)$ satisfies the
fragmentation equation with constant $p(s)$ and hence by (\ref{l21}) we have
\begin{equation*}
\int_{0}^{\infty }x^{\frac{2}{\gamma }-1}\left\vert g(x,t)\right\vert
^{2}dx\leq Ce^{-2\sigma t}\int_{0}^{\infty }x^{\frac{2}{\gamma }%
-1}\left\vert g(x,0)\right\vert ^{2}dx,
\end{equation*}%
with $\sigma =1$ if $\gamma >1$ and $\sigma =\gamma $ if $\gamma \leq 1$.
Since%
\begin{equation*}
\frac{C^{-1}}{(1+\left\vert z\right\vert ^{2})^{\frac{\nu }{2}}}\leq \frac{1%
}{\left\vert F(z)\right\vert }\leq \frac{C}{(1+\left\vert z\right\vert
^{2})^{\frac{\nu }{2}}},
\end{equation*}%
we conclude, for $\nu \leq 0$%
\begin{equation*}
\int_{0}^{\infty }x^{\frac{2}{\gamma }-1}\left\vert D_{\left\vert \nu
\right\vert }(u(x,t))\right\vert ^{2}dx\leq Ce^{-2\sigma t}\int_{0}^{\infty
}x^{\frac{2}{\gamma }-1}\left\vert D_{\left\vert \nu \right\vert
}(u(x,0)\right\vert ^{2}dx,
\end{equation*}%
where%
\begin{equation*}
\widetilde{D_{\left\vert \nu \right\vert }u}(z,t)=(1+\left\vert z\right\vert
^{2})^{\frac{\left\vert \nu \right\vert }{2}}\widetilde{u}(z,t),
\end{equation*}%
and for $\nu >0$, by (\ref{estd}) with $i=\left[ \nu \right] $ derivatives%
\begin{equation*}
\int_{0}^{\infty }x^{\frac{2}{\gamma }-1}\left\vert D_{\nu -\left[ \nu %
\right] }(u(x,t))\right\vert ^{2}dx\leq Ce^{-2\sigma t}\int_{0}^{\infty }x^{%
\frac{2}{\gamma }-1}\left\vert D_{\nu -\left[ \nu \right] }(u(x,0))\right%
\vert ^{2}dx.
\end{equation*}

\section{Extensions and limitations}

In this final section we will discuss on the possible extensions of our
theory and its limitations. Note that equation (\ref{eqnMellin}), for a
function $p(s)$ that is not strictly a polynomial but an arbitrary finite
linear combination of powers:%
\begin{equation*}
p(s)=\sum_{i=0}^{N}a_{i}s^{\mu _{i}},
\end{equation*}%
such that $\mu _{0}=0$, $\Re (\mu _{i})>0$ for $i>0$. If $\mu _{s}$ is
complex, then its complex conjugate is also among $\left\{ \mu _{i}\right\} $
($p(s)$ must be real). Then,%
\begin{equation*}
P(z)=\int_{0}^{1}s^{z-1}p(s^{\frac{1}{\gamma }})ds=\sum_{i=0}^{N}\frac{a_{i}%
}{z+\mu _{i}/\gamma },
\end{equation*}%
and we can write equation (\ref{eqnMellin}) in the form
\begin{equation}
\widetilde{u}_{t}(z,t)-\gamma \left( z-\frac{2}{\gamma }\right) \left[
\widetilde{u}(z,t)-\frac{1}{z}\frac{P_{N}(z)}{Q_{N}(z)}\widetilde{u}(z+1,t)%
\right] =0,
\end{equation}%
where, now%
\begin{eqnarray*}
P_{N}(z) &=&\prod_{i=1}^{N}(z-z_{i}), \\
Q_{N}(z) &=&\prod_{i=1}^{N}\left( z+\frac{\mu _{i}}{\gamma }\right)
\end{eqnarray*}%
with $z_{i}$ the roots of a $N$-th degree polynomial. It is simple to verify
that $\sum_{i=1}^{N}(i/\gamma +z_{i})=\sum_{i=0}^{N}\frac{a_{i}}{\gamma }-%
\frac{2}{\gamma }=\frac{p(1)-2}{\gamma }$. We can write%
\begin{equation*}
\widetilde{U}(z,t)=\frac{\prod_{i=1}^{N}\Gamma (z+\frac{\mu _{i}}{\gamma })}{%
\prod_{i=1}^{N}\Gamma (z-z_{i})}\Psi (z,t),
\end{equation*}%
so that $\Psi (z,t)$ solves%
\begin{equation}
\Psi _{t}(z,t)-\gamma \left( z-\frac{2}{\gamma }\right) \left[ \Psi (z,t)-%
\frac{1}{z}\Psi (z+1,t)\right] =0,  \label{eqnpsi}
\end{equation}%
for which Theorem \ref{th2} applies. We can now estimate as in (\ref{ff}):%
\begin{equation}
\left\vert \frac{\prod_{i=1}^{N}\Gamma (z+\frac{\mu _{i}}{\gamma })}{%
\prod_{i=1}^{N}\Gamma (z-z_{i})}\right\vert \leq C(1+\lambda ^{2})^{\frac{%
p(1)-2}{2}},  \label{estim}
\end{equation}%
and prove Theorem \ref{th3} in an identical fashion.

Another extension is to the degenerate case when the polynomial $p(s)$ is
zero at $s=0$. Assume $p(s)=\sum_{i=q}^{N}a_{i}s^{q}$. Then, when $N=q$, the
eigenvalue equation (\ref{eigenvalue}) is%
\begin{equation*}
\left( -\mu -\gamma z+2\right) \widetilde{U}(z)=\left( \frac{q+2}{z+\frac{q}{%
\gamma }}-\gamma \right) \widetilde{U}(z+1),
\end{equation*}%
with solutions%
\begin{equation*}
\widetilde{U}(z)=\frac{\Gamma \left( -z+\frac{2}{\gamma }+1\right) }{\Gamma
\left( -z+\frac{2-\mu }{\gamma }+1\right) }\Gamma \left( z+\frac{q}{\gamma }%
\right) ,
\end{equation*}%
and eigenfunctions%
\begin{equation*}
U_{n}(x)=n!x^{\frac{q}{\gamma }}L_{n}^{\left( \frac{q+2}{\gamma }-n\right)
}(x)e^{-x},
\end{equation*}%
so that Theorem \ref{th1} holds with an extra $x^{-2q}$ factor in the first
condition in (\ref{cond2}) and
\begin{eqnarray*}
u(x,t) &=&\sum_{n=1}^{\infty }a_{n}e^{-n\gamma t}n!x^{q}L_{n}^{\left( \frac{%
q+2}{\gamma }-n\right) }(x^{\gamma })e^{-x^{\gamma }} \\
a_{n} &=&\sum_{j=n}^{\infty }\frac{j!}{(n!)^{2}(j-n)!}\frac{\gamma }{\Gamma (%
\left[ \frac{2}{\gamma }\right] +1)}\int_{0}^{\infty }y^{\gamma
+1+q}L_{j}^{\left( \left[ \frac{q+2}{\gamma }\right] \right) }(y^{\gamma
})v_{0}(y)dy.
\end{eqnarray*}%
Theorems \ref{th3} and \ref{th4} also hold in this case.

A natural question that arises is whether our estimates generalize to more
general $p(s)$ such as Dirac measures. The answer is negative in general
since, although the function $P(z)$ might have a discrete set of poles and
even be meromorphic, the presence of infinite poles may create essential
singularities of $P(z)$ at infinity. Note at this respect that a meromorphic
function in $\widehat{%
\mathbb{C}
}$ (the complex plane counting also possible poles at infinity, i.e. poles
of $P(1/z)$) must be rational and hence contain a finite number of poles. As
a simple illustration, lets take%
\begin{equation*}
p(s)=2\delta (s-1/2),
\end{equation*}%
so that%
\begin{equation*}
P(z)=2\gamma \int_{0}^{1}s^{z\gamma -1}\delta (s-1/2)ds=\gamma 2^{2-z\gamma
},
\end{equation*}%
which does not grow polynomically fast at infinity (no poles at infinity).
So it is not possible to write $\frac{1}{\gamma }P(z)-1$ as a rational
function. We can still formally write
\begin{equation*}
\widetilde{u}(z,t)=F(z)\Psi (z,t),
\end{equation*}%
with $\Psi (z,t)$ satisfying (\ref{eqnpsi}) and $F(z)$ such that%
\begin{equation*}
F(z)-\frac{z}{2-\gamma z}\left( \frac{1}{\gamma }P(z)-1\right) F(z+1)=0,
\end{equation*}%
so that $F(z)$ may contain essential singularities at infinity. In that case
an estimate like (\ref{estim}) is not verified, and we cannot expect Theorem %
\ref{th3} to hold in ordinary Sobolev spaces but in rather restrictive
topologies. This can also be seen, for instance, in $F(z)$ satisfying%
\begin{equation*}
F(z)-\frac{1}{\Gamma (z)}F(z+1)=0
\end{equation*}%
which is the so-called Barnes G-function, behaving asymptotically as $%
F(z+1)\sim e^{z^{2}\log z+O(z^{2})}$ as $z\rightarrow \infty $.

Nevertheless, for real analytic $p(s)$, under certain conditions, the theory
may be adapted. Let $p(s)$ be of the form
\begin{equation}
p(s)=\sum_{i=0}^{\infty }a_{i}s^{i},  \label{psana}
\end{equation}%
(with $a_{0}>0$) so that $P(z)=\sum_{i=0}^{\infty }\frac{a_{i}}{z+i/\gamma }$
and assume $K(z)=\gamma \left( \frac{1}{\gamma }P(z)-1\right) $ satisfies (%
\ref{assumpt}). Consider the problem%
\begin{equation}
F(z)-\frac{\gamma z}{2-\gamma z}\left( \frac{1}{\gamma }P(z)-1\right)
F(z+1)=0.  \label{eqefe}
\end{equation}%
We can prove the following Lemma:

\begin{lemma}
\bigskip Let $p(s)$ given by (\ref{psana}) with
\begin{equation}
\left\vert a_{j}\right\vert \leq \frac{L}{j^{1+\delta }},  \label{aj}
\end{equation}%
for some $L,\delta >0$. Then there is a solution to (\ref{eqefe}) given by%
\begin{equation*}
F(z)=\frac{\prod_{j=1}^{\infty }\Gamma (z+\frac{j}{\gamma })}{%
\prod_{j=1}^{\infty }\Gamma (z-z_{j})},
\end{equation*}%
for certain $\left\{ z_{j}\right\} $ and such that, for $\lambda \in
\mathbb{R}
$, there exists a constant $C$, independent $\lambda $, such that%
\begin{equation*}
C^{-1}(1+\lambda ^{2})^{\frac{p(1)-2}{2\gamma }}\leq \left\vert F\left(
i\lambda +\frac{1}{2}\right) \right\vert \leq C(1+\lambda ^{2})^{\frac{p(1)-2%
}{2\gamma }}.
\end{equation*}%
Moreover, one can write $F(z)=F_{1}(z)F_{2}(z)$ with $F_{1}(z)$ a finite
product of quotients of gamma functions and $F_{2}(z)$ satisfying%
\begin{equation}
C^{-1}\leq \left\vert F_{2}\left( i\lambda +\frac{1}{2}+M\right) \right\vert
\leq C,  \label{estimma}
\end{equation}%
for any $M\geq 0$ and $C$ independent of $M$ and $\lambda $.
\end{lemma}

\textbf{Proof. }In order to solve (\ref{eqefe}) we are interested first in
finding the roots of
\begin{equation}
K(z)=\gamma \left( \sum_{i=0}^{\infty }\frac{a_{i}}{\gamma z+i}-1\right) ,
\label{kk}
\end{equation}%
located at $\left\vert z\right\vert \gg 1$. We will show that they are real
and close to the poles at $\left\{ -j/\gamma \right\} $. We look at the
roots in the form%
\begin{equation*}
z=-\frac{j}{\gamma }+\widetilde{z},
\end{equation*}%
so that%
\begin{equation*}
\gamma \left( \sum_{i=0}^{N}\frac{a_{i}}{i-j+\gamma \widetilde{z}}-1\right) =%
\frac{a_{j}}{\widetilde{z}}-\gamma +\sum_{i\neq j}^{N}\frac{a_{i}}{%
i-j+\gamma \widetilde{z}},
\end{equation*}%
and%
\begin{equation*}
K(z)=0\Longrightarrow \gamma -\frac{a_{j}}{\widetilde{z}}=\sum_{i\neq j}^{N}%
\frac{a_{i}}{i-j+\gamma \widetilde{z}}.
\end{equation*}%
We further write%
\begin{equation*}
\widetilde{z}=\frac{a_{j}(1+x)}{\gamma },
\end{equation*}%
so that $x$ satisfies the equation%
\begin{equation}
\frac{\gamma x}{(1+x)}=\sum_{i\neq j}^{N}\frac{a_{i}}{i-j+a_{j}(1+x)}.
\label{eqx}
\end{equation}%
We estimate the right hand side assuming $\left\vert x\right\vert <\frac{1}{%
10}$ and $j$ sufficiently large so that
\begin{eqnarray*}
\left\vert \sum_{i\neq j}^{N}\frac{a_{i}}{i-j+a_{j}(1+x)}\right\vert &\leq
&M\sum_{i\neq j}^{N}\frac{1}{i^{1+\delta }\left\vert i-j\right\vert -\frac{1%
}{10}} \\
&\leq &M\left( \sum_{i=1}^{j-1}\frac{1}{(i+1)^{1+\delta }\left\vert
i-j\right\vert }+\sum_{i=j+1}^{N}\frac{1}{(i+1)^{1+\delta }\left\vert
i-j\right\vert }\right) \\
&\leq &C\frac{\log j}{j},
\end{eqnarray*}%
and consequently, a standard fixed point argument shows existence of a
solution to (\ref{eqx}) for arbitrary $N$ sufficiently large with
\begin{equation*}
\left\vert x\right\vert <C\frac{\log j}{j},
\end{equation*}%
uniformly in $N$. Hence, zeros of $K(z)$ are at a distance smaller than $%
O(j^{-1-\delta })$ from poles for $j>j_{0}$.

We consider now
\begin{equation*}
G(z)=\frac{\prod_{j=1}^{\infty }(\gamma z-\gamma z_{j})}{\prod_{j=1}^{\infty
}(\gamma z+j)},
\end{equation*}%
that can be written as%
\begin{equation*}
G(z)=\left[ \prod_{j=1}^{j_{0}-1}\frac{(\gamma z+j)}{(\gamma z-\gamma z_{j})}%
\right] \frac{(\gamma z-\gamma \widetilde{z_{j_{0}}})}{(\gamma z-\gamma
z_{j_{0}})}\left[ \prod_{j=j_{0}}^{\infty }\frac{(\gamma z+j)}{(\gamma
z-\gamma \widetilde{z_{j}})}\right] ,
\end{equation*}%
with $\widetilde{z_{j}}=z_{j}$ for $j>j_{0}$ and $\widetilde{z_{j_{0}}}$
such that $\sum_{j_{0}}^{\infty }(j+\gamma \widetilde{z_{j}})=0$. Then%
\begin{equation*}
F(z)=\frac{\prod_{j=1}^{\infty }\Gamma (z+\frac{j}{\gamma })}{%
\prod_{j=1}^{\infty }\Gamma (z-z_{j})},
\end{equation*}%
and we can factor%
\begin{equation*}
F(z)=F_{1}(z)F_{2}(z),
\end{equation*}%
with
\begin{eqnarray*}
F_{1}(z) &=&\left[ \prod_{j=1}^{j_{0}-1}\frac{\Gamma (z+\frac{j}{\gamma })}{%
\Gamma (z-z_{j})}\right] \frac{\Gamma (z-\widetilde{z_{j_{0}}})}{\Gamma
(z-z_{j_{0}})} \\
F_{2}(z) &=&\prod_{j=j_{0}}^{\infty }\frac{\Gamma (z+\frac{j}{\gamma })}{%
\Gamma (z-\widetilde{z_{j}})}.
\end{eqnarray*}%
As in previous sections, we estimate%
\begin{equation*}
C^{-1}(1+\lambda ^{2})^{\frac{p(1)-2}{2}}\leq \left\vert F_{1}(i\lambda +%
\frac{1}{2})\right\vert \leq C(1+\lambda ^{2})^{\frac{p(1)-2}{2}}.
\end{equation*}%
We note now that
\begin{equation*}
F_{2}(z)=\prod_{j=j_{0}}^{\infty }\frac{\Gamma (z+\frac{j}{\gamma })}{\Gamma
\left( \left( z+\frac{j}{\gamma }\right) +\varepsilon _{j}\right) },
\end{equation*}%
with $\varepsilon _{j}=-\left( \widetilde{z_{j}}+\frac{j}{\gamma }\right) $
and $\left\vert \varepsilon _{j}\right\vert =O(j^{-1-\delta })$. In order to
estimate $F_{2}(i\lambda +1/2)$ we estimate the following quotient for $\Re
(z)>0$ and $\left\vert z\right\vert >1$:%
\begin{equation*}
g(z)=\frac{\Gamma (z)}{\Gamma (z+\varepsilon )}.
\end{equation*}%
Using Binet's first formula (cf. \cite{E2})%
\begin{equation*}
\ln \Gamma (z)=\left( z-\frac{1}{2}\right) \ln z-z+\frac{1}{2}\ln (2\pi
)+\int_{0}^{\infty }\left( \frac{1}{2}-\frac{1}{t}+\frac{1}{e^{t}-1}\right)
\frac{e^{-tz}}{t}dt,
\end{equation*}%
we compute%
\begin{eqnarray*}
\ln \Gamma (z)-\ln \Gamma (z+\varepsilon ) &=&\left( z-\frac{1}{2}\right)
\ln z-\left( z+\varepsilon -\frac{1}{2}\right) \ln (z+\varepsilon ) \\
&&+\varepsilon +\int_{0}^{\infty }\left( \frac{1}{2}-\frac{1}{t}+\frac{1}{%
e^{t}-1}\right) \frac{e^{-tz}(1-e^{-t\varepsilon })}{t}dt \\
&=&-\varepsilon \ln (z+\varepsilon )-\left( z-\frac{1}{2}\right) \ln (1+%
\frac{\varepsilon }{z})+\varepsilon \\
&&+\int_{0}^{\infty }\frac{(t-2)(e^{t}-1)+2t}{2t^{2}(e^{t}-1)}%
e^{-tz}(1-e^{-t\varepsilon })dt \\
&=&-\varepsilon \ln z-O\left( \frac{\varepsilon }{z}\right)
+\int_{0}^{\infty }\frac{(t-2)(e^{t}-1)+2t}{2t^{2}(e^{t}-1)}%
e^{-tz}(1-e^{-t\varepsilon })dt,
\end{eqnarray*}%
and using%
\begin{equation*}
\frac{(t-2)(e^{t}-1)+2t}{2t^{2}(e^{t}-1)}=\frac{1}{12}+O(t),
\end{equation*}%
so that%
\begin{equation*}
\frac{(t-2)(e^{t}-1)+2t}{2t^{2}(e^{t}-1)}\leq Ce^{-t}.
\end{equation*}%
For $\Re (z)>0$
\begin{equation*}
\left\vert \int_{0}^{\infty }\frac{(t-2)(e^{t}-1)+2t}{2t^{2}(e^{t}-1)}%
e^{-tz}(1-e^{-t\varepsilon })dt\right\vert \leq C\int_{0}^{\infty
}e^{-t}(e^{\left\vert \varepsilon \right\vert t}-1)e^{-\Re (z)t}dt\leq
C\left\vert \varepsilon \right\vert ,
\end{equation*}%
where $C$ is independent of $\left\vert z\right\vert $ (for $\left\vert
z\right\vert $ sufficiently large) and then,
\begin{equation*}
\left\vert -\left( z+\frac{1}{2}\right) \ln (1+\frac{\varepsilon }{z}%
)+\varepsilon +\int_{0}^{\infty }\frac{(t-2)(e^{t}-1)+2t}{2t^{2}(e^{t}-1)}%
e^{-tz}(1-e^{-t\varepsilon })dt\right\vert \leq C\left\vert \varepsilon
\right\vert ,
\end{equation*}%
where $C$ is independent of $\left\vert z\right\vert $, so that%
\begin{equation*}
g(z)=z^{-\varepsilon }e^{h(z;\varepsilon )},
\end{equation*}%
where $\left\vert e^{h(z;\varepsilon )}\right\vert \leq e^{C\left\vert
\varepsilon \right\vert }.$ We have then%
\begin{eqnarray*}
\prod_{j=j_{0}}^{\infty }\frac{\Gamma (z+\frac{j}{\gamma })}{\Gamma (z-%
\widetilde{z_{j}})} &=&\prod_{j=j_{0}}^{\infty }(z+j/\gamma )^{-\varepsilon
_{j}}e^{h(z+j/\gamma ;\varepsilon _{j})} \\
&=&e^{-\sum_{j_{0}}^{\infty }\varepsilon _{j}\log (z+j/\gamma
)}e^{\sum_{j_{0}}^{\infty }h(z+j/\gamma ;\varepsilon _{j})},
\end{eqnarray*}%
with $\varepsilon _{j}=-(j/\gamma +\widetilde{z_{j}})$. Using now $%
\sum_{j=j_{0}}^{\infty }\varepsilon _{j}=0$ we can write, for any $M\geq 0$,
\begin{eqnarray*}
&&\sum \varepsilon _{j}\ln (j/\gamma +i\lambda +M+1/2)-\sum \varepsilon
_{j}\ln (1+i\lambda +M) \\
&=&\sum \varepsilon _{j}\ln (1+\frac{j/\gamma +1/2}{1+i\lambda +M}),
\end{eqnarray*}%
and since%
\begin{equation*}
\sum \left\vert \varepsilon _{j}\ln \left( 1+\frac{j/\gamma +1/2}{1+i\lambda
+M}\right) \right\vert \leq C\sum \left\vert \varepsilon _{j}\right\vert
\log j\leq C,
\end{equation*}%
with $C$ independent of $M$ and $\lambda $, and%
\begin{equation*}
\left\vert e^{\sum_{j_{0}}^{\infty }h(i\lambda +j/\gamma +1/2;\varepsilon
_{j})}\right\vert \leq e^{C\sum \left\vert \varepsilon _{j}\right\vert },
\end{equation*}%
we conclude%
\begin{equation*}
\left\vert \prod_{j=j_{0}}^{\infty }\frac{\Gamma (i\lambda +1/2+M+j/\gamma )%
}{\Gamma (i\lambda +1/2+M-\widetilde{z_{j}})}\right\vert \leq C.
\end{equation*}%
Following identical steps for $F_{2}^{-1}$ we can estimate%
\begin{equation*}
\left\vert \prod_{j=j_{0}}^{\infty }\frac{\Gamma (i\lambda +1/2+M-\widetilde{%
z_{j}})}{\Gamma (i\lambda +1/2+M+j/\gamma )}\right\vert \leq C,
\end{equation*}%
and this completes the proof of the Lemma.

Using estimate (\ref{estimma}) and dealing with $F_{1}(z)$ as we did with $%
F(z)$ in the proof of Theorem \ref{th3} one can show Theorem \ref{th3} also
for $p(s)$ given by (\ref{psana}) with (\ref{aj}). The same applies to
Theorem \ref{th4}.

\section{Appendix A: Mellin transforms}

By simple integration one can compute the following Mellin transforms (cf.
\cite{E}):

1.-%
\begin{eqnarray}
f(x) &=&e^{-x},  \notag \\
F(z) &=&\Gamma (z),  \label{p1}
\end{eqnarray}

\bigskip (formula (1) in 6.3 of \cite{E})

2.-%
\begin{eqnarray}
f(x) &=&x^{a}e^{-x},  \notag \\
F(z) &=&\Gamma (z+a),\ \Re (a)>0,  \label{p2}
\end{eqnarray}

(formula (1) in 6.3 and (3) of 6.1 of \cite{E})

3.-%
\begin{eqnarray}
f(x) &=&\Gamma (a)M(a,b,-x),\ \Re (a)>0,  \notag \\
F(z) &=&\frac{\Gamma (a-z)}{\Gamma (b-z)}\Gamma (z),  \label{p3}
\end{eqnarray}%
where $M(a,b,-x)$ is a confluent hypergeometric (or Kummer's) function. \
This is a direct consequence of the Barnes type integral 13.2.9 in \cite{A}.
Of interest for us is the following relation for the asymptotic behavior of
Kummer's functions (cf. formula 13.5.1 and Kummer's transformation 13.1.27
in \cite{A}):
\begin{equation}
M(a,b,z)\sim \Gamma (b)\left( \frac{e^{z}z^{a-b}}{\Gamma (b)}+\frac{(-z)^{-a}%
}{\Gamma (b-a)}\right) ,\text{ as }\left\vert z\right\vert \rightarrow
\infty ,\ -\frac{3\pi }{2}<\arg z\leq \frac{\pi }{2}.  \label{asy}
\end{equation}

4.- For $\ \Re (\nu )>0$,
\begin{eqnarray}
f(x) &=&\left\{
\begin{array}{c}
\frac{1}{\Gamma (\nu )}(1-x)^{\nu -1},\ x\leq 1 \\
0,\ x>1%
\end{array}%
\right. ,  \notag \\
F(z) &=&\frac{\Gamma (z)}{\Gamma (z+\nu )},  \label{p4}
\end{eqnarray}%
(formula (20) of 7.3 in \cite{E}). The so-called Riemann-Liouville
fractional integral (for $0<\nu <1$) defined as
\begin{equation*}
I_{\nu }(u)=\frac{1}{\Gamma (\nu )}\int_{0}^{x}(x-y)^{\nu -1}u(y)dy,
\end{equation*}%
has as Mellin transform
\begin{equation*}
\widetilde{I_{\nu }(u)}(z)=\frac{\Gamma (z)}{\Gamma (z+\nu )}\widetilde{u}%
(z+\nu ).
\end{equation*}

5.- For $n\in
\mathbb{N}
$, $\alpha \in
\mathbb{R}
$%
\begin{equation}
f(x)=L_{n}^{(\alpha )}(x)e^{-x},  \label{p51}
\end{equation}%
where $L_{n}^{(\alpha )}(x)$ is the generalized Laguerre polynomial defined
by means of Rodrigues formula (22.11.6 in \cite{A}):%
\begin{equation*}
L_{n}^{(\alpha )}(x)=\frac{e^{x}x^{-\alpha }}{n!}\frac{d^{n}}{dx^{n}}%
(x^{n+\alpha }e^{-x}).
\end{equation*}%
Then%
\begin{equation}
F(z)=\frac{(-1)^{n}}{n!}(z-\alpha -1)(z-\alpha -2)...(z-\alpha -n)\Gamma (z).
\label{p52}
\end{equation}

\section{Appendix B: Estimates on Gamma functions}

\bigskip In this appendix we will obtain some estimates, involving Gamma
functions, that are used in previous sections. We start with the following
relation (formula 6.1.40 in \cite{A}):
\begin{equation*}
\log \Gamma (z+h)\sim \left( z+h-\frac{1}{2}\right) \log z-z+\frac{1}{2}\log
(2\pi )+O(z^{-1}),
\end{equation*}%
valid for finite $h\in
\mathbb{R}
$ and $\left\vert z\right\vert \rightarrow \infty $. Then, for $a,b\in
\mathbb{C}
$,%
\begin{equation*}
\log \Gamma (z+a)-\log \Gamma (z+b)\sim a\log z-b\log z+O(z^{-1}),
\end{equation*}%
and hence%
\begin{equation}
f(z)\equiv \frac{\Gamma (z+a)}{\Gamma (z+b)}\sim z^{a-b},  \label{estf}
\end{equation}%
as $\left\vert z\right\vert \rightarrow \infty $. If
\begin{equation*}
a\neq 0,-1,-2,...
\end{equation*}%
then $f(z)$ has no poles, $\left\vert f(z)\right\vert $ is bounded in any
compact set and using (\ref{estf}) we have then%
\begin{equation}
\left\vert \frac{\Gamma (i\lambda +a)}{\Gamma (i\lambda +b)}\right\vert \leq
C(1+\lambda ^{2})^{\frac{a-b}{2}}.  \label{estf1}
\end{equation}%
Analogously, for $a_{i},b_{i}\in
\mathbb{C}
$, we have%
\begin{equation*}
f(z)\equiv \frac{\prod\limits_{i=1}^{N}\Gamma (z+a_{i})}{\prod%
\limits_{i=1}^{N}\Gamma (z+b_{i})}\sim z^{\sum_{i=1}^{N}(a_{i}-b_{i})},
\end{equation*}%
as $\left\vert z\right\vert \rightarrow \infty $. If
\begin{equation*}
a_{i}\neq 0,-1,-2,...,
\end{equation*}%
then%
\begin{equation}
\left\vert \frac{\prod\limits_{i=1}^{N}\Gamma (i\lambda +a_{i})}{%
\prod\limits_{i=1}^{N}\Gamma (i\lambda +b_{i})}\right\vert \leq C(1+\lambda
^{2})^{\frac{1}{2}\sum_{i=1}^{N}(a_{i}-b_{i})}.  \label{estfmm}
\end{equation}%
If, in addition,
\begin{equation*}
b_{i}\neq 0,-1,-2,...,
\end{equation*}%
then%
\begin{equation*}
\left\vert \frac{\prod\limits_{i=1}^{N}\Gamma (i\lambda +a_{i})}{%
\prod\limits_{i=1}^{N}\Gamma (i\lambda +b_{i})}\right\vert \geq
C^{-1}(1+\lambda ^{2})^{\frac{1}{2}\sum_{i=1}^{N}(a_{i}-b_{i})}.
\end{equation*}

We also need to estimate%
\begin{equation*}
f(z)\equiv \frac{\Gamma (i\lambda +a+M+1)}{\Gamma (i\lambda +b+M+1)},
\end{equation*}%
for large values of $M$. We can write
\begin{eqnarray*}
&&\frac{\Gamma (i\lambda +a+M+1)}{\Gamma (i\lambda +b+M+1)} \\
&=&\frac{(i\lambda +a+M)(i\lambda +a+M-1)...(i\lambda +a+M-\left[ M\right] )%
}{(i\lambda +b+M)(i\lambda +b+M-1)...(i\lambda +b+M-\left[ M\right] )}\frac{%
\Gamma (i\lambda +a+\nu )}{\Gamma (i\lambda +b+\nu )},
\end{eqnarray*}%
and define%
\begin{equation*}
g(\lambda )=\left\vert \frac{i\lambda +a+s}{i\lambda +b+s}\right\vert ^{2}=%
\frac{\lambda ^{2}+(a+s)^{2}}{\lambda ^{2}+(b+s)^{2}}.
\end{equation*}%
If $a\leq b$ then $g(\lambda )$ is increasing and then $g(\lambda )\leq 1$
for any $s\geq 0$. Hence%
\begin{equation}
\left\vert \frac{\Gamma (i\lambda +a+M+1)}{\Gamma (i\lambda +b+M+1)}%
\right\vert \leq \left\vert \frac{\Gamma (i\lambda +a+\nu )}{\Gamma
(i\lambda +b+\nu )}\right\vert \leq C(1+\lambda ^{2})^{\frac{a-b}{2}},
\label{estfm}
\end{equation}%
where $C$ is independent of $M$. If, on the other hand, $a>b$, $g(\lambda )$
is decreasing and then $g(\lambda )\leq \frac{(a+s)^{2}}{(b+s)^{2}}$%
\begin{eqnarray*}
&&\left\vert \frac{(i\lambda +a+M)(i\lambda +a+M-1)...(i\lambda +a+M-\left[ M%
\right] )}{(i\lambda +b+M)(i\lambda +b+M-1)...(i\lambda +b+M-\left[ M\right]
)}\right\vert \\
&\leq &\frac{\Gamma (b+\nu )}{\Gamma (a+\nu )}\frac{\Gamma (a+M+1)}{\Gamma
(b+M+1)}\leq CM^{a-b}.
\end{eqnarray*}%
We estimate now%
\begin{equation*}
f(z)=\prod_{i=1}^{N}\frac{\Gamma (i\lambda +a_{i}+M+1)}{\Gamma (i\lambda
+b_{i}+M+1)},
\end{equation*}%
and define%
\begin{eqnarray*}
g(\lambda ) &=&\left\vert \prod \frac{i\lambda +a_{i}+s}{i\lambda +b_{i}+s}%
\right\vert ^{2}=\frac{\left( \lambda ^{2}+(a_{1}+s)^{2}\right) \left(
\lambda ^{2}+(a_{2}+s)^{2}\right) ...\left( \lambda
^{2}+(a_{N}+s)^{2}\right) }{\left( \lambda ^{2}+(b_{1}+s)^{2}\right) \left(
\lambda ^{2}+(b_{2}+s)^{2}\right) ...\left( \lambda
^{2}+(b_{N}+s)^{2}\right) } \\
&=&\frac{\left( \left( \frac{\lambda }{s}\right) ^{2}+(\frac{a_{1}}{s}%
+1)^{2}\right) \left( \left( \frac{\lambda }{s}\right) ^{2}+(\frac{a_{2}}{s}%
+1)^{2}\right) ...\left( \left( \frac{\lambda }{s}\right) ^{2}+(\frac{a_{N}}{%
s}+1)^{2}\right) }{\left( \left( \frac{\lambda }{s}\right) ^{2}+(\frac{b_{1}%
}{s}+1)^{2}\right) \left( \left( \frac{\lambda }{s}\right) ^{2}+(\frac{b_{2}%
}{s}+1)^{2}\right) ...\left( \left( \frac{\lambda }{s}\right) ^{2}+(\frac{%
b_{N}}{s}+1)^{2}\right) } \\
&=&\frac{\prod \left[ 1+\frac{\frac{2a_{i}}{s}+\left( \frac{a_{i}}{s}\right)
^{2}}{1+\left( \frac{\lambda }{s}\right) ^{2}}\right] }{\prod \left[ 1+\frac{%
\frac{2b_{i}}{s}+\left( \frac{b_{i}}{s}\right) ^{2}}{1+\left( \frac{\lambda
}{s}\right) ^{2}}\right] }.
\end{eqnarray*}%
For $s\gg \left\vert a_{i}\right\vert ,\left\vert b_{i}\right\vert $ for any
$i$ we have%
\begin{eqnarray*}
\log g(\lambda ) &=&\sum \log \left[ 1+\frac{\frac{2a_{i}}{s}+\left( \frac{%
a_{i}}{s}\right) ^{2}}{1+\left( \frac{\lambda }{s}\right) ^{2}}\right] -\sum
\log \left[ 1+\frac{\frac{2b_{i}}{s}+\left( \frac{b_{i}}{s}\right) ^{2}}{%
1+\left( \frac{\lambda }{s}\right) ^{2}}\right] \\
&\sim &2\frac{\sum (a_{i}-b_{i})}{1+\left( \frac{\lambda }{s}\right) ^{2}}.
\end{eqnarray*}%
If $\sum (a_{i}-b_{i})\leq 0$ then $g(\lambda )$ is increasing and
\begin{equation}
\left\vert \prod \frac{\Gamma (i\lambda +a_{i}+M+1)}{\Gamma (i\lambda
+b_{i}+M+1)}\right\vert \leq C\left\vert \prod \frac{\Gamma (i\lambda
+a_{i}+\nu )}{\Gamma (i\lambda +b_{i}+\nu )}\right\vert \leq C(1+\lambda
^{2})^{\frac{\sum (a_{i}-b_{i})}{2}},  \label{estfff}
\end{equation}%
where $C$ is independent of $M$. If $\sum (a_{i}-b_{i})>0$ then $g(\lambda )$
is decreasing and%
\begin{equation}
\left\vert \frac{\Gamma (i\lambda +a_{i}+M+1)}{\Gamma (i\lambda +b_{i}+M+1)}%
\right\vert \leq CM^{\sum (a_{i}-b_{i})}(1+\lambda ^{2})^{\frac{\sum
(a_{i}-b_{i})}{2}}.  \label{estff}
\end{equation}

\section{Appendix C: Explicit representation formula}

The explicit representation formula (\ref{fxt}) can also be directly
verified by inserting it into (\ref{eqnt}). Using (\ref{fxt}) we can compute
\begin{eqnarray*}
u_{t} &=&-\frac{1}{\Gamma (1-\frac{2}{\gamma })}e^{-t}x^{-\frac{1}{\gamma }%
}\int_{0}^{1}e^{-xw(1-e^{-\gamma t})}F(e^{-\gamma t}xw)w^{\frac{1}{\gamma }%
}(1-w)^{-\frac{2}{\gamma }}dw \\
&&-\frac{\gamma }{\Gamma (1-\frac{2}{\gamma })}e^{-\gamma t}e^{-t}x^{1-\frac{%
1}{\gamma }}\int_{0}^{1}e^{-xw(1-e^{-\gamma t})}F^{\prime }(e^{-\gamma
t}xw)w^{1+\frac{1}{\gamma }}(1-w)^{-\frac{2}{\gamma }}dw \\
&&-\frac{\gamma }{\Gamma (1-\frac{2}{\gamma })}e^{-t}e^{-\gamma t}x^{1-\frac{%
1}{\gamma }}\int_{0}^{1}e^{-xw(1-e^{-\gamma t})}F(e^{-\gamma t}xw)w^{\frac{1%
}{\gamma }+1}(1-w)^{-\frac{2}{\gamma }}dw,
\end{eqnarray*}%
and%
\begin{eqnarray*}
&&\gamma x^{1-\frac{1}{\gamma }}\frac{\partial }{\partial x}(x^{\frac{1}{%
\gamma }}u) \\
&=&-\frac{\gamma }{\Gamma (1-\frac{2}{\gamma })}x^{1-\frac{1}{\gamma }%
}e^{-t}\int_{0}^{1}e^{-xw(1-e^{-\gamma t})}(1-e^{-\gamma t})F(e^{-\gamma
t}xw)w^{\frac{1}{\gamma }+1}(1-w)^{-\frac{2}{\gamma }}dw \\
&&+\frac{\gamma }{\Gamma (1-\frac{2}{\gamma })}x^{1-\frac{1}{\gamma }%
}e^{-t}e^{-\gamma t}\int_{0}^{1}e^{-xw(1-e^{-\gamma t})}F^{\prime
}(e^{-\gamma t}xw))w^{\frac{1}{\gamma }+1}(1-w)^{-\frac{2}{\gamma }}dw,
\end{eqnarray*}%
so that%
\begin{eqnarray}
&&u_{t}+\gamma x^{1-\frac{1}{\gamma }}\frac{\partial }{\partial x}(x^{\frac{1%
}{\gamma }}u)+u  \notag \\
&=&-\frac{1}{\Gamma (1-\frac{2}{\gamma })}\gamma x^{1-\frac{1}{\gamma }%
}e^{-t}\int_{0}^{1}e^{-xw(1-e^{-\gamma t})}F(e^{-\gamma t}xw)w^{\frac{1}{%
\gamma }+1}(1-w)^{-\frac{2}{\gamma }}dw.  \label{fff}
\end{eqnarray}%
The right hand side of (\ref{fff}), that we call $R\left[ u\right] $, must
equal the fragmentation term (\ref{eqnt}):%
\begin{equation*}
\mathbb{F}(u)=\gamma \left( \frac{2}{\gamma }\int_{x}^{\infty
}u(y)dy-xu(z)\right) .
\end{equation*}%
We will show it by using Mellin transform. First, The Mellin transform of $R%
\left[ u\right] $ is
\begin{equation*}
R\left[ u\right] (z)=-\frac{\gamma }{\Gamma (1-\frac{2}{\gamma })}\int x^{z-%
\frac{1}{\gamma }}\left( e^{-t}\int_{0}^{1}e^{-xw(1-e^{-\gamma
t})}F(e^{-\gamma t}xw)w^{\frac{1}{\gamma }+1}(1-w)^{-\frac{2}{\gamma }%
}dw\right) dx
\end{equation*}%
\begin{eqnarray*}
&=&-\frac{\gamma }{\Gamma (1-\frac{2}{\gamma })}\int (xw)^{z-\frac{1}{\gamma
}+1}\left( e^{-t}\int_{0}^{1}e^{-xw(1-e^{-\gamma t})}F(e^{-\gamma t}xw)w^{-z+%
\frac{2}{\gamma }}(1-w)^{-\frac{2}{\gamma }}dw\right) d(xw) \\
&=&-\gamma \left( \int (v^{z-\frac{1}{\gamma }+1}e^{-v(1-e^{-\gamma
t})}F(e^{-\gamma t}v)dv\right) \frac{\Gamma (-z+\frac{2}{\gamma }+1)}{\Gamma
(-z+2)} \\
&=&\frac{\gamma }{\pi }\left( \int (v^{z-\frac{1}{\gamma }%
+1}e^{-v(1-e^{-\gamma t})}F(e^{-\gamma t}v)dv\right) \Gamma (-z+\frac{2}{%
\gamma }+1)\Gamma (z-1)\sin (\pi z),
\end{eqnarray*}%
and since the Mellin transform of $u$ at $z+1$ is
\begin{eqnarray*}
\widetilde{u}(z+1) &=&\frac{1}{\Gamma (1-\frac{2}{\gamma })}\int e^{-t}x^{z-%
\frac{1}{\gamma }}\left( \int_{0}^{1}e^{-xw(1-e^{-\gamma t})}F(e^{-\gamma
t}xw)w^{\frac{1}{\gamma }}(1-w)^{-\frac{2}{\gamma }}dw\right) dx \\
&=&\frac{1}{\Gamma (1-\frac{2}{\gamma })}\int e^{-t}(xw)^{z-\frac{1}{\gamma }%
}\int_{0}^{1}e^{-xw(1-e^{-\gamma t})}F(e^{-\gamma t}xw)w^{\frac{2}{\gamma }%
-z-1}(1-w)^{-\frac{2}{\gamma }}dw \\
&=&\left( \int (v^{z-\frac{1}{\gamma }}e^{-v(1-e^{-\gamma t})}F(e^{-\gamma
t}v)dv\right) \frac{\Gamma (-z+\frac{2}{\gamma })}{\Gamma (1-z)} \\
&=&\frac{1}{\pi }\left( \int (v^{z-\frac{1}{\gamma }}e^{-v(1-e^{-\gamma
t})}F(e^{-\gamma t}v)dv\right) \Gamma (-z+\frac{2}{\gamma })\Gamma (z)\sin
(\pi z),
\end{eqnarray*}%
we conclude, using%
\begin{eqnarray*}
\frac{1}{\Gamma (1-z)} &=&\frac{1}{\pi }\Gamma (z)\sin (\pi z), \\
\frac{1}{\Gamma (2-z)} &=&\frac{1}{\pi }\Gamma (z-1)\sin (\pi (z-1))=-\frac{1%
}{\pi }\Gamma (z-1)\sin (\pi z),
\end{eqnarray*}%
that%
\begin{equation*}
R\left[ u\right] (z)=\gamma \widetilde{u}(z+1)\frac{\frac{2}{\gamma }-z}{z},
\end{equation*}%
which is exactly the expression for the fragmentation term in Mellin
transform representation.

\begin{acknowledgement}
We thank Jos\'{e} A. Ca\~{n}izo for fruitful conversations and suggestions
on the topic of this article.
\end{acknowledgement}


\begin{thebibliography}{99}
\bibitem{Ajmone09} G.~{Ajmone Marsan}. \newblock New paradigms towards the
modelling of complex systems in behavioral economics.
\newblock {\em Math.
Comput. Modelling}, 50:584--597, 2009.

\bibitem{A} M. Abramowitz and I. A. Stegun, Handbook of Mathematical
Functions. With. Formulas, Graphs, and Mathematical Tables, National Bureau
of Standards Applied Mathematics Series, No. 55, 1970.

\bibitem{AjmBellomoEgidi08} G.~{Ajmone Marsan}, N.~Bellomo, and M.~Egidi. %
\newblock Towards a mathematical theory of complex socio-economical systems
by functional subsystems representation.
\newblock {\em Kinet. Relat.
Models}, 1:249--278, 2008.

\bibitem{Aldous99} David~J. Aldous. \newblock Deterministic and stochastic
models for coalescence (aggregation and coagulation): a review of the
mean-field theory for probabilists. \newblock {\em Bernoulli}, 5(1):3--48,
1999.

\bibitem{B1} J.~Banasiak, Analytic Methods for Coagulation-Fragmentation
Models, Volume I (Chapman \& Hall/CRC Monographs and Research Notes in
Mathematics) 2019.

\bibitem{B2} J.~Banasiak, Analytic Methods for Coagulation-Fragmentation
Models, Volume II (Chapman \& Hall/CRC Monographs and Research Notes in
Mathematics) 2019.

\bibitem{AB} J.~Banasiak and L.~Arlotti.
\newblock {\em Perturbations of
positive semigroups with applications}. \newblock Springer Monographs in
Mathematics. Springer-Verlag London, 2006.

\bibitem{BalaCaGabr13} Daniel Balagu{\'e}, Jos{\'e}~A. Ca{\~n}izo, and
Pierre Gabriel. \newblock Fine asymptotics of profiles and relaxation to
equilibrium for growth-fragmentation equations with variable drift rates. %
\newblock {\em Kinet. Relat. Models}, 6(2):219--243, 2013.

\bibitem{BellomoEtAll2009} N.~Bellomo, C.~Bianca, and M.~Delitala. \newblock %
Complexity analysis and mathematical tools towards the modelling of living
systems. \newblock {\em Phys. Life Rev.}, 6:144--175, 2009.

\bibitem{BellouBian10} A.~Bellouquid and C.~Bianca. \newblock Modelling
aggregation-fragmentation phenomena from kinetic to macroscopic scales. %
\newblock {\em Math. Comput. Modelling}, 52(5-6):802--813, 2010.

\bibitem{Bertoin06} Jean Bertoin.
\newblock {\em Random fragmentation and
coagulation processes}, volume 102 of \emph{Cambridge Studies in Advanced
Mathematics}. \newblock Cambridge University Press, Cambridge, 2006.

\bibitem{BreschiFontelos} G. Breschi and M. A. Fontelos, A note on the
self-similar solutions to the spontaneous fragmentation equation, Proc. R.
Soc. A 473 (2017), 20160740.

\bibitem{CaceCaMisc11} Mar{\'{\i}}a~J. C{\'{a}}ceres, Jos{\'{e}}~A. Ca{\~{n}}%
izo, and St{\'{e}}phane Mischler. \newblock Rate of convergence to an
asymptotic profile for the self-similar fragmentation and
growth-fragmentation equations. \newblock {\em J. Math. Pures Appl. (9)},
96(4):334--362, 2011.

\bibitem{C} L. Carlitz, Some generating functions for Laguerre polynomials,
Duke Math. J., 35 (1968), 825--827.

\bibitem{ChengRed90} Z.~Cheng and S.~Redner. \newblock Kinetics of
fragmentation. \newblock {\em J. Phys. A}, 23(7):1233--1258, 1990.

\bibitem{ClaLash99} Christophe Clanet and Juan~C. Lasheras. \newblock %
Transition from dripping to jetting. \newblock {\em J. Fluid Mech.},
383:307--326, 1999.

\bibitem{Drake72review} R.~Drake. \newblock A general mathematical survey of
the coagulation equation. \newblock In George~M. Hidy and James~Rush Brock,
editors, \emph{Topics in current aerosol research}, volume Part 2, pages
201--376, 1972.

\bibitem{EastwArmiLash04} C.~D. Eastwood, L.~Armi, and J.~C. Lasheras. %
\newblock The breakup of immiscible fluids in turbulent flows. \newblock%
\emph{Journal of Fluid Mechanics}, 502:309--333, Mar 2004. \newblock n/a.

\bibitem{EF} J. Eggers, M. A. Fontelos, The role of self-similarity in
singularities of partial differential equations, Nonlinearity 22, R1 (2009).

\bibitem{E} A. Erdelyi, Tables on integral transforms, vol. 1, McGraw-Hill,
New York,1954.

\bibitem{E2} A. Erd\'{e}lyi, W. Magnus, F. Oberhettinger and F. G. Tricomi,
Higher Transcendental Functions, Vol. 1. New York: Krieger, 1981.

\bibitem{EscoMischRic05} M.~Escobedo, S.~Mischler, and M.~Rodriguez~Ricard. %
\newblock On self-similarity and stationary problem for fragmentation and
coagulation models.
\newblock {\em Ann. Inst. H. Poincar\'e Anal. Non
Lin\'eaire}, 22(1):99--125, 2005.

\bibitem{FasanoEtAll06} A.~Fasano, F.~Rosso, and A.~Mancini. \newblock %
Implementation of a fragmentation--coagulation-scattering model for the
dynamics of stirred liquid--liquid dispersions. \newblock {\em Phys. D},
222:141--158, 2006.

\bibitem{GabrielSalvarani} P. Gabriel and F. Salvarani, Exponential
relaxation to self-similarity for the superquadratic fragmentation equation,
Applied Mathematics Letters, 2014, 27, pp.74-78.

\bibitem{Kostoglou01b} M.~Kostoglou and A.J. Karabelas. \newblock On the
breakage problem with a homogeneous erosion type kernel.
\newblock {\em
Journal of Physics A: Mathematical and General}, 34(8):1725--1740, 2001. %
\newblock cited By (since 1996)12.

\bibitem{LasherasEtAll02} J.C. Lasheras, C.~Eastwood, C.~Mart\'{\i}nez-Baz{%
\'{a}}n, and J.L. Monta{\~{n}}{\'{e}}s. \newblock A review of statistical
models for the break-up of an immiscible fluid immersed into a fully
developed turbulent flow.
\newblock {\em International Journal of Multiphase
Flow}, 28(2):247 -- 278, 2002.

\bibitem{LaurMisch04} Philippe Lauren{\c{c}}ot and St{\'e}phane Mischler. %
\newblock On coalescence equations and related models. \newblock In \emph{%
Modeling and computational methods for kinetic equations}, Model. Simul.
Sci. Eng. Technol., pages 321--356. Birkh\"auser Boston, Boston, MA, 2004.

\bibitem{LaurBen09} Philippe Lauren{\c{c}}ot and Benoit Perthame. \newblock %
Exponential decay for the growth-fragmentation/cell-division equation. %
\newblock {\em Commun. Math. Sci.}, 7(2):503--510, 2009.

\bibitem{Leyv03} Fran{\c{c}}ois Leyvraz. \newblock Scaling theory and
exactly solved models in the kinetics of irreversible aggregation.
\newblock
\emph{Phys. Rep.}, 383:95--212, 2003.

\bibitem{Michel06b} P.~Michel. \newblock Optimal proliferation rate in a
cell division model. \newblock {\em Math. Model. Nat. Phenom.}, 1(2):23--44,
2006.

\bibitem{Michel06} Philippe Michel. \newblock Existence of a solution to the
cell division eigenproblem. \newblock {\em Math. Models Methods Appl. Sci.},
16(7, suppl.):1125--1153, 2006.

\bibitem{MS} S. Mischler, J. Scher, Spectral analysis of semigroups and
growth-fragmentation equations, Annales de l'Institut Henri Poincar\'{e} C,
Analyse non lin\'{e}aire 33-3, 849-898, 2016.

\bibitem{ParisKa} R.B. Paris and D.~Kaminski.
\newblock {\em Asymptotics and
Mellin-Barnes integrals}, volume~85 of \emph{\ Encyclopedia of Mathematics
and its applications}. \newblock Cambridge University Press, Cambridge, 2001.

\bibitem{PM} R. P\'{e}rez-Marco, On the definition of higher Gamma
functions, arXiv:2101.01291.

\bibitem{Treat97b} Richard~P. Treat. \newblock On the similarity solution of
the fragmentation equation. \newblock {\em J. Phys. A}, 30(7):2519--2543,
1997.

\bibitem{Treat97a} Richard~P. Treat. \newblock Similarity solution for
fragmentation with volume change. \newblock {\em J. Phys. A},
30(21):7639--7658, 1997.

\bibitem{Wattis06} J.A.D. Wattis. \newblock An introduction to mathematical
models of coagulation--fragmentation processes: a discrete deterministic
mean-field approach. \newblock {\em Phys. D}, 222:1--20, 2006.

\bibitem{ZhaoEtAll09} Z.~Zhao, A.~Kirou, B.~Ruszczycki, and N.F. Johnson. %
\newblock Dynamical clustering as a generator of complex system dynamics. %
\newblock {\em Math. Models Methods Appl. Sci.}, 19 (Suppl.):1539--1565,
2009.

\bibitem{Ziff91} Robert~M. Ziff. \newblock New solutions to the
fragmentation equation. \newblock {\em J. Phys. A}, 24(12):2821--2828, 1991.

\bibitem{ZiffMcGrady85} Robert~M. Ziff and E.~D. McGrady. \newblock The
kinetics of cluster fragmentation and depolymerisation.
\newblock {\em J.
Phys. A}, 18(15):3027--3037, 1985.
\end{thebibliography}
\end{document}